\documentclass[a4paper]{article}
\newcommand{\affil}[1]{$^{\rm #1}$}
\usepackage[authoryear]{natbib}
\bibpunct{(}{)}{;}{a}{}{,}
\usepackage{graphicx}
\usepackage{amssymb} 
\usepackage{lscape}
\date{} 
\newcommand{\Msun}{{\rm M}_{\odot}}

\newcommand{\iso}[2]{\hbox{${}^{#1}{\rm #2}$}}
\newcommand{\Cratio}{\iso{12}C/\iso{13}C}

\title{\large\bf\flushleft  Stellar Models and Yields of Asymptotic Giant Branch Stars}
\author{\parbox{\textwidth}{\flushleft
\vspace{-0.5cm}
{\it Amanda Karakas\affil{A,C} and John C. Lattanzio\affil{B}}\\
\vspace{0.4cm}
{\small \affil{A}\,Research School of Astronomy \& Astrophysics,
Mount Stromlo Observatory, Cotter Road Weston Creek, ACT 2611, Australia}\\
{\small \affil{B}\,Centre for Stellar and Planetary Astrophysics, 
Monash University, PO Box 28M, Clayton VIC 3800, Australia}\\
{\small \affil{C}\,Email: akarakas@mso.anu.edu.au}}}
\begin{document}
%
\begin{changemargin}{.8cm}{.5cm}
\begin{minipage}{.9\textwidth}
\vspace{-1cm}
\maketitle
%
%
\small{\bf Abstract:}
We present stellar yields calculated from detailed models of low
and intermediate-mass asymptotic giant branch (AGB) stars. We evolve
models with a range of mass from 1 to 6$\Msun$, and initial
metallicities from solar to $1/200^{\rm th}$ of the solar metallicity.
Each model was evolved from the zero age main sequence to near the end
of the thermally-pulsing AGB (TP-AGB) phase, and through all intermediate phases
including the core He-flash for stars initially less massive 
than 2.5$\Msun$.
For each mass and metallicity, we provide tables containing structural 
details of the stellar models during the TP-AGB phase, and tables of 
the stellar yields for 74 species from hydrogen through to sulphur, 
and for a small number of iron-group nuclei. All tables are available
for download.  Our results have many applications 
including use in population synthesis studies and the chemical
evolution of galaxies and stellar systems, and for comparison to
the composition of AGB and post-AGB stars and planetary nebulae.
\medskip\\{\bf Keywords:} stars: AGB and post-AGB --- abundances 
--- ISM: abundances --- planetary nebulae: general

\medskip
\medskip
\end{minipage}
\end{changemargin}
\small

\section{Introduction}

Stars with initial masses in the range $\sim 0.8$ to 8$\Msun$, depending
on the initial metallicity $Z$, will pass through the thermally-pulsing 
asymptotic giant branch (TP-AGB) phase before ending their lives as white dwarfs.
The structure and evolution of low and intermediate mass stars prior to and
during the AGB has been previously discussed in some detail
\citep[see][ and references therein]{busso99,herwig05}.
An AGB star is characterized by two nuclear burning shells,
one burning helium (He) above a degenerate carbon-oxygen core and
another burning hydrogen (H), below a deep convective envelope.  
In-between lies the intershell region composed mostly of \iso{4}He.
The helium-burning shell is thermally unstable and flashes or pulses 
every 10$^{4}$~years or so.
After the occurrence of a thermal pulse (TP), mixing episodes may occur that
bring the products of nuclear burning from deep inside
the star to the stellar surface. These mixing events, or third dredge-up (TDU)
episodes, bring the products of partial He-burning (mostly \iso{4}He and 
\iso{12}C) into the envelope.  The TDU is the mechanism for turning (single) 
stars into carbon stars, where the C/O ratio exceeds unity in the surface layers.
Following dredge-up the H-shell is re-ignited and the star enters a phase
of quiescent H-burning known as the interpulse phase. The TP--AGB is 
defined as the phase from the first TP to the time when the
star ejects its outer envelope, terminating the AGB phase.

In the most massive AGB models ($M \gtrsim 4\Msun$, depending on the
metallicity) hot bottom burning (HBB) can occur when the base of the 
convective envelope dips into the top of the H-shell  resulting in a 
thin layer hot enough to sustain 
proton-capture nucleosynthesis. The convective turn-over time in the 
envelope is $\sim 1$ year that ensures that the whole envelope is exposed
to the hot region a few thousand times per interpulse period.
There is observational evidence that HBB
occurs in massive AGB stars \citep{wood83,smith89,mcsaveney07} 
in the Large and Small Magellanic Clouds (LMC and SMC). 
HBB converts \iso{12}C into \iso{14}N and if the TDU is operating, 
HBB will prevent the atmosphere from becoming carbon rich
\citep*{boothroyd93}. The copious amounts of
\iso{14}N produced in this case will be primary\footnote{produced 
from the H and \iso{4}He initially present in the star.} 
owing to the primary \iso{12}C being dredged from the He-shell. 
\citet{frost98a} found intermediate-mass AGB evolution 
to be sensitive to the initial composition and the mass-loss law 
used in the calculation, with models becoming C-rich near the
tip of the AGB if TDU continues after HBB has been shut off.
The occurrence and duration of HBB depends upon the initial metallicity, 
mass loss and convection \citep{ventura05a} where the minimum mass for 
HBB is pushed to lower mass in lower metallicity models 
\citep{frost98a}.

Owing to fact that calculating a TP-AGB model is a computationally intensive
task, synthetic AGB models, which use fitting formulae to model the evolution
quickly, have proved to be a successful approach to population syntheses studies 
that require large stellar populations. Historically this 
approach was validated by the fact that the stellar luminosity on
the AGB is nearly a linear function of the H-exhausted core mass 
\citep[e.g.][]{paczynski75, renzini81, groen93} although we know now that this 
relation breaks down for stars undergoing HBB \citep{bloecker91,lattanzio92}.
Synthetic AGB models have successfully been used to model AGB 
populations \citep{groen93,marigo96} and compute stellar yields 
\citep{vandenhoek97,marigo01,izzard04b}.
Many of the parameterizations used in synthetic evolution studies are derived 
from detailed stellar models, such as the growth of the H-exhausted 
core with time, and as such are only accurate over the range in mass and 
metallicity of the stellar models they are based upon.
An example is provided by \citet{vandenhoek97} who compute AGB yields
for initial masses between 0.9 and 8$\Msun$ whereas the \citet{boothroyd88c}
interpulse-period-core mass relation they use was only derived for stars
with initial masses between 1 and 3$\Msun$.  What affect this has on the
yields is unclear since this relation will affect the number of TPs
during the TP-AGB phase and hence the level of chemical
enrichment.  Dredge-up and HBB affect these relationships but owing 
to our lack of  understanding about these phenomena, there are still 
considerable uncertainties that affect the detailed computations.  
Recent improvements in computer power mean that  grids of detailed AGB 
models can now be produced in a reasonable time 
\citep{ventura02,karakas02,herwig04b} although producing yields from 
$N \gtrsim 20$ AGB stars for any given metallicity range is still  
challenging. For this reason synthetic models are still preferred
for some applications.

In an effort to try and quantify the contribution of AGB stars to 
the production of elements in the Galaxy we have calculated grids 
of detailed AGB models covering a substantial range 
in initial mass and metallicity.  The aim of this paper is to
provide the results of these grids in a format suitable for 
many applications. We first present the AGB stellar structure in
a format useful for comparison to other detailed models, or for use
in synthetic AGB algorithms.  Second, we provide the stellar yields.
These could be useful for different applications including 
chemical evolution studies, for comparison to the composition of 
planetary nebulae (PNe) and pre-solar grains.  The data are 
provided in a tabulated format available for download 
via the internet, where the appropriate reference for these 
tables is this paper.

These models were previously discussed in detail in \citet{karakasThesis} 
and in a number of other studies that help to highlight the importance 
of these models for both detailed and synthetic AGB calculations. 
These studies include \citet{karakas02}, \citet{karakas03b}, \citet{izzard04b},
\citet{lugaro04}, \citet{karakas06a}, \citet{lugaro07}, \citet{izzard07} 
and \citet{karakas07a}. These later studies used the structure models 
presented here and changed the reaction rates and/or initial abundances 
to study particular problems e.g. the effect of reaction rate 
uncertainties on the production of \iso{19}F in AGB stars \citep{lugaro04}.

We begin with a brief description of the numerical method.

\section{The Numerical Method}

The numerical method we use has been previously described in detail
\citep[e.g.][]{lugaro04,karakas06a}. Here we summarize the main points relevant for
this study. We compute the structure first and then perform detailed
nucleosynthesis calculations. 

\subsection{The Stellar Structure models}

The stellar structure models were calculated with the Monash version of the 
Mount Stromlo Stellar Structure Program; see \citet{frost96} and references 
therein for details.  Mass loss on the first giant branch is included using the 
Reimer's mass-loss prescription \citep[][ hereafter R75]{reimers75} 
with the parameter $\eta = 0.4$; on the AGB we use the formulation 
given by \citet[][ hereafter VW93]{vw93} in all models unless 
indicated otherwise. We calculate two models (5$\Msun$, $Z=0.02$ and $Z=0.0001$) 
using Reimer's mass loss on the AGB with the parameter $\eta = 3.5$.
We used the solar abundances of \citet{anders89} for the $Z = 0.02$ models.
For the $Z = 0.008$ and $0.004$ models we used C, N, and O abundances 
from \citet{russell92} that are appropriate for stars in the 
Large and Small Magellanic Clouds where many AGB studies are performed.
We assumed scaled-solar abundances for the $Z  = 0.0001$ models.

All models were calculated from the zero-age main sequence to near the end of
the TP-AGB phase. For models with initial masses below 2.5$\Msun$ we also 
evolve the models through the core He-flash. While this is an approximation to the 
true 3D nature of the core flash \citep{dearborn06}, it at least allows us to model 
the entire evolution of the star self-consistently without resorting to using 
zero-age horizontal branch models or other techniques to get around this 
event \citep[see for e.g.][]{stancliffe04a}.

The occurrence of the TDU and HBB depend critically
upon the convection model used \citep{frost96,mowlavi99a,ventura05a} and 
the method for determining convective borders.  We use the standard
mixing-length theory (MLT) for convective regions with a mixing-length
parameter $\alpha = l / H_P = 1.75$, and determine the border by applying the
Schwarzschild criterion. Hence we do not include convective overshoot,
in the usual sense.  Instead we search for a neutral border to the
convective zone in the manner described by \citet{lattanzio86}. 
This method has been shown to increase the efficiency of the TDU 
compared to models that strictly use the Schwarzschild criterion
\citep{frost96}.

We will not provide a detailed discussion of the 
uncertainties that affect our models but we refer the interested 
reader to \citet{frost96}, \citet{busso99}, \citet{goriely00},
\citet{ventura05a,ventura05b} and \citet{herwig05}
noting that this is a non-exhaustive list of publications 
dedicated to discussing uncertainties in the modelling of AGB
stars.

\subsection{Post-processing Nucleosynthesis models}

We performed detailed nucleosynthesis calculations using a post-processing
code that includes 74 species and time-dependent diffusive mixing in all
convective zones \citep{cannon93}. The details of the
nucleosynthesis network are outlined in \citet{lugaro04} but we remind the
reader that we include 59 light nuclei and 14 iron-group species.
We also add the fictional particle $g$ to count the number of neutron
captures occurring beyond \iso{62}Ni  \citep{lattanzio96,lugaro04}.

The bulk of the 506 reaction rates are from the {\tt REACLIB} data tables 
\citep{thielemann86}, based on the 1991 updated version.  Some of the proton, 
$\alpha$ and neutron capture reaction rates have been updated according to the 
latest experimental results, see \citet{lugaro04} for details.

\section{The Stellar Models}  \label{modelresults}

The mass grid used for each metallicity is shown in Table~\ref{models} 
where we note if the model experienced the core He-flash, TDU and/or HBB. 
For each model AGB star we present structural information in tabulated 
form and these tables are available for download from the website
{\tt http://www.mso.anu.edu.au/\~{}akarakas/model\_data/}.  
Samples are given in Tables~\ref{example1} and~\ref{example2}.
Data is provided as a function of the TP number as a proxy for time.
Each table contains the pulse number, and for that pulse we include 
the core mass $M_{\rm core}$, the maximum extent of the pulse-driven 
convection region $M_{\rm csh}$, the maximum duration of the convective 
pocket $t_{\rm csh}$, the amount of matter dredged into the envelope 
$\Delta M_{\rm dredge}$, the parameters $\lambda$ and $\lambda_{dup}$ 
(see below), and the maximum temperature in the He-shell $T_{\rm Heshell}$
(measured from the middle of the He-shell which is hotter than the bottom). 
We include the maximum temperature at the base of the convective envelope 
$T_{\rm bce}$, and the maximum temperature in the H-shell $T_{\rm Hshell}$;
these were computed over the previous interpulse period. 
We then include the interpulse period, the total mass (measured at 
the current TP) $M_{\rm tot}$, the maximum  radiated luminosity 
during the previous interpulse period Max~L, and the maximum 
He-luminosity during the current TP Max~LHe. All units are in 
solar units with the exception of temperatures, that are in kelvin, 
and all times which are in years.

\begin{table}[t]
\begin{center}
\caption{Grids of stellar masses for each $Z$, noting if the models 
experience the core He-flash (CHe), the third dredge-up (TDU), and 
hot bottom burning (HBB).}\label{models}
\begin{tabular}{cllll}
\hline mass &  $Z = 0.02$ & $Z = 0.008$ & $Z = 0.004$ & $Z=10^{-4}$ \\
\hline 
\hline
 1.0 & CHe & CHe & CHe & -- \\
1.25 & CHe & CHe & CHe & CHe,TDU \\
1.5  & CHe & CHe & CHe,TDU & -- \\
1.75 & CHe & CHe,TDU & CHe,TDU & CHe,TDU \\
1.9  & CHe & CHe,TDU & CHe,TDU & --  \\
2.0  & CHe & --  & --  & TDU \\
2.1  & --  & CHe,TDU & -- & -- \\
2.25 & CHe,TDU & TDU & TDU & TDU \\
2.5  & TDU & TDU & TDU & TDU \\
3.0  & TDU & TDU & TDU & TDU,HBB \\
3.5  & TDU & TDU & TDU & TDU,HBB \\
4.0  & TDU & TDU,HBB & TDU,HBB & TDU, HBB \\
5.0  & TDU,HBB & TDUHBB & TDU,HBB & TDU,HBB \\
6.0  & TDU,HBB & TDU,HBB & TDU,HBB & TDU,HBB \\
6.5  & TDU,HBB & --  & --  & --  \\
7.0  & --  & --  & --  & TDU,HBB$^{a}$ \\
\hline
\hline
\end{tabular}
\medskip\\
$a$ This model goes through central C-burning and is a super-AGB star.\\
\end{center}
\end{table}

The TDU efficiency parameter, $\lambda$, is usually defined 
according to $\lambda = \Delta M_{\rm dredge}/\Delta M_{\rm h}$, 
where $\Delta M_{\rm h}$  is the amount by which the core mass 
has grown between the present and previous TPs. 
In our tables we provide $\lambda$ along with the parameter 
$\lambda_{\rm dup}$, defined by \citet{goriely00} to be 
$\Delta M_{\rm dredge}/M_{\rm csh}$.  This quantity is useful for 
comparison to the results of \citet{goriely00} and also because it 
provides a measure of the amount of pulse-driven convective mass mixed 
into the outer envelope.  In the study by \citet{goriely00} they 
assumed a constant value of $\lambda_{\rm dup} = 0.1$ whereas we 
observe this parameter to increased as a function of TP number
in models that experience the TDU.

We had originally aimed to evolve each model $\Delta \log T_{\rm eff} \sim 0.3$ 
off the AGB track, to the point where the entire outer envelope 
of the model had been lost. Owing to convergence difficulties near 
the tip of the AGB this was only achieved in one case: the 1$\Msun$, 
$Z=0.02$ model where mass loss removed the entire outer envelope, 
leaving the model with a final remnant mass of 0.573$\Msun$ and on 
the white-dwarf cooling track with a final $\log T_{\rm eff} = 4.118$. 
The low-mass models ($M < 2\Msun$) lost most of their outer envelopes,
even if they did not leave the AGB track completely according to the 
definition above.  For most of these models the remaining envelope 
mass was small, much less than $0.1\Msun$ and we would not expect 
further TPs on the AGB.
For the more massive models the remaining envelope mass was large 
enough that further TPs would probably occur if we could 
evolve the models further along the AGB track. In all cases
 however the luminosity-driven superwind phase had started and 
the envelope mass was being reduced at a rapid rate, at a few 
$\times \, 10^{-5} \, \Msun$ year$^{-1}$.  HBB had ceased 
to operate in all models; this occurred once the envelope mass was 
reduced below about $\sim 1.5 \Msun$.  In \S\ref{sectionyields}
we discuss how we estimated the number of remaining TPs and
the contribution of these pulses to the chemical enrichment
of the model star.

The convergence difficulties mostly occurred during the start of the final 
TDU phase, after the TP has peaked in He-luminosity.  The reasons for the 
convergence problems are unclear but they are not associated with small
values of the gas pressure to total pressure, $\beta$, since previous 
dredge-up episodes occur without problem for similar values of $\beta$. 
The convergence problems are sometimes associated with a rapid drop in 
the temperature at the base of the convective envelope and an increase 
in the stellar radius. Higher metallicity models, at a given initial
mass, experience convergence difficulties earlier (i.e. at a larger 
envelope mass) than models of lower metallicity. This may indicate that 
the input physics, notably the opacities, are related to the problem. 
A thorough investigation into the cause of the convergence difficulties 
is required.

Two models require a special mention and these are the 6 and 7$\Msun$,
$Z = 0.0001$ models. The 6$\Msun$ model was evolved through more
than 100 TPs but by the time we ended the computation, little envelope
mass had been removed, the superwind phase had not been
reached and HBB was actively changing the composition of the outer
envelope. For this model we estimate that another 100 could possibly
occur when using the \citet{vw93} mass-loss prescription. We do not
attempt to estimate the contribution from the remaining TPs owing to
the complication caused by active HBB; see \citet{izzard06} for
the difficulties associated with this. The 7$\Msun$ model went through
degenerate central carbon burning before ascending the TP-AGB
phase. For this model we only computed 12 TPs before we ended the
computation. We point out that it  is unclear if this model 
would end its life as a massive ONe white dwarf or it would reach the
Chandrasekhar limiting mass and explode as a low-mass Type II 
supernova. At the end of the computation, the final core mass was 
1.126$\Msun$ and with inefficient TDU and a low mass-loss rate we
suspect it would explode as a low-mass Type II supernova 
but further study is required to determine the final fate of this model.

In Tables~\ref{example1} and ~\ref{example2} we present the results for the 
3$\Msun$, $Z = 0.02$ model, to show an example of the tables 
available. Note that these data are  given as one table for each 
mass and $Z$ in the on-line version.  For the case shown the TDU begins 
at the 10$^{\rm th}$ TP when the core mass reaches 0.62$\Msun$.  
Adding up the $\Delta M_{\rm dredge}$ for each pulse gives a total of 
0.0787$\Msun$ core material that is added to the envelope over the 
course of the  TP-AGB phase.   This model does not experience HBB and 
that can be seen by inspection of Table~\ref{example2} where the maximum 
temperature at the base of the convective envelope during the interpulse 
does not exceed 10$^{7}\,$K.  Other interesting comparisons can be made 
by considering the evolution of the interpulse period as a function of 
the core mass. From the data presented in Table~\ref{example2} we 
observe an initial increase followed by slow decrease.
We can compare to the core mass--interpulse  period relations given 
in  \citet{boothroyd88c} or \citet{wagenhuber98}
at the final core mass of the model, 0.68$\Msun$. The first relation
gives a value of 3.47 $\times 10^{4}$ years and the second 
$3.73 \times 10^{4}$ years.
Both of these values are significantly smaller than the value that
comes out of the detailed computation of $5.47 \times 10^{4}$ years. 
Hence synthetic AGB simulations using either of these relations 
would result in many more TPs during the TP-AGB than we predict 
and hence more TDU episodes and a greater level of chemical enrichment.

\subsection{The Third Dredge-up and HBB}

Owing mostly to the operation of TDU episodes AGB stars are 
important for enriching the composition of the interstellar medium 
(ISM).  Observational evidence suggests that the TDU occurs in 
stars as low as $\sim 1.2\Msun$ at
solar metallicity and in the LMC \citep{frogel90,wallerstein98} and at
$\sim 1\Msun$ in the SMC \citep{frogel90}. An examination of the tables 
presented above show that the minimum mass for the TDU at solar 
metallicity is 2.25$\Msun$, and 1.5$\Msun$
at the metallicities appropriate for the LMC ($Z = 0.008$)
and SMC ($Z = 0.004$). In \citet{karakas02} we discussed
in detail how these models do not predict enough TDU at small
enough core masses to account for the observations. The
recent analysis of infra-red  data by \citet{guandalini06} 
suggests that the minimum mass of stars experiencing
the TDU has previously been greatly under-estimated, owing to 
incorrect estimates of bolometric magnitudes.  However if the
study of \citet{stancliffe05a} is anything to go on
there is still a requirement for more efficient TDU in 
low-mass low-metallicity stars than we predict.

Massive AGB stars that experience HBB are one of the favoured
sites of the abundance anomalies observed in galactic globular
cluster stars \citep[][]{yong03a,gratton04}, even if problems
persist in comparisons between the model predictions
and the observed abundance trends 
\citep{denissenkov03b,fenner04,ventura05b,karakas06b}. 
Observations of stars experiencing HBB have mostly come from the
LMC and SMC \citep{wood83,plez93}, with some observations
of stars in our Galaxy \citep{garcia06,garcia07}. 
These observations indicate that most luminous AGB stars 
have an O-rich composition (where C/O $< 1$) with a low 
\Cratio~ratio, and that many are also rich in Li \citep{smith90b},
and in elements produced by the slow-neutron capture process.
Observations by \citet{mcsaveney07} have shown that two 
bright O-rich AGB stars are rich in nitrogen and depleted 
in carbon. These observations are the first observational 
confirmation of the long-predicted production of primary 
nitrogen by the combination of TDU and HBB in 
intermediate-mass AGB stars. 

\section{The Stellar Yields} \label{sectionyields}

The stellar yields can be downloaded from
the website: {\tt http://www.mso.anu.edu.au/\~{}akarakas/stellar\_yields/}.  
We present the stellar yields in two ways. The first method
presents the yields and abundances in the wind integrated over the
entire stellar lifetime and these are suitable for use in e.g. galactic
chemical evolution studies. An example is given in Table~\ref{yield1}.
Each of the tables contain the following: 1) the nuclear species $i$,
2) the atomic mass A($i$), 3) the nett stellar yield, defined below, 
4) the amount of species $i$ in the wind lost from the star 
mass$(i)_{\rm lost} \, (\Msun)$, 5) the amount of $i$ that would 
initially have been present in the wind mass$(i)_{0}\, (\Msun)$.
The quantity mass$(i)_{0}$ is simply the mass expelled 
during the stellar lifetime multiplied by the initial abundance.
We next include 6) the average mass fraction of $i$ in the wind
$\langle X(i)\rangle$, 7) the initial mass fraction $X_{0}(i)$ and
8) the production factor $f$ defined by 
$\log [ \langle X(i)\rangle / X_{0}(i) ]$. Note that in Table~\ref{yield1}
the production factor  has simply been labelled $f$ whereas the 
definition is provided in the header of the on-line tables.
This last quantity $f$ is useful for comparison to authors who
present their results in a similar way  
\citep[see Table~4 from ][]{herwig04b} as well as for comparison to
derived abundances from stars, where for e.g. an $f \approx 0.2$ 
would indicate an increase of that species by 0.2~dex at the surface.

The definition of the yield that we use is given by the following 
expression
\begin{equation}
 M_{\rm i} = \int_{0}^{\tau} \left[ X(i) - X_{0} (k)\right] \frac{d M}{dt} dt,
\label{eq:netyield}
\end{equation}
where $M_{i}$ is the yield of species $i$ (in solar masses),
$dM/dt$ is the current mass-loss rate, $X(i)$ and $X_{0} (i)$ refer
to the current and initial mass fraction of species $i$, and $\tau$ is the
total lifetime of the stellar model. The yield can be negative, in the 
case where the element is destroyed, and positive if it is produced.
We also present the total amount of $i$ (in $\Msun$) expelled into
the ISM, noting that this value is always positive.

It was previously discussed that some of the models do not lose 
their entire outer envelopes during the TP--AGB evolution owing to 
convergence difficulties near the tip of the AGB. This may be 
important for the stellar yields because, depending on the amount 
of remaining envelope, these models may experience further TPs and
TDU episodes. Even one more TDU episode could significantly alter the
surface abundances at this stage because the envelope mass is small
leading to less dilution. The affect on the stellar yields is probably
smaller because this quantity is integrated over the entire stellar
lifetime.  

For each model we estimate the number of remaining TPs; this is
shown in Table~\ref{missedTPs} along with the TDU efficiency
parameter from the last computed TP, $\lambda_{\rm f}$.
Models not included in this table have small enough envelope 
masses that we estimate that no further TPs would take place. 
For the models in Table~\ref{missedTPs} we present two sets of 
stellar yields, one with no contribution from these extra pulses; 
this is where we remove the envelope at its given composition 
and assume that $\lambda = 0$ during remaining pulses.  
The other set of yields were computed using a constant value 
of $\lambda$ for the remaining TPs. This latter set are 
our ``standard'' set and we recommend these be used.  
The set with $\lambda = 0$ are useful for
exploring model uncertainties.

From inspection of Table~\ref{missedTPs} an example of 
a model where remaining TPs might affect
the yields is the 6$\Msun$, $Z = 0.02$ model where we estimate that
5 more TPs could occur. The magnitude of the change to the surface 
composition from these final TPs depends on one big unknown:  
The behaviour of $\lambda$ with decreasing envelope mass. 
Detailed stellar models suggests that $\lambda$ decreases with 
decreasing envelope mass \citep[][ and data presented 
here]{straniero97,karakas02} but \citet{stancliffe07} find results 
to the contrary and instead suggest that there is no reduction in 
$\lambda$, at least for the 1.5$\Msun$ models discussed 
in that paper.

In Table~\ref{yield1} we present a set of stellar yields for the 
3$\Msun$, $Z = 0.02$ model.
From Table~\ref{missedTPs} we estimate that one more TP would 
occur for this model and the yields presented here include this 
final TP.  In Table~\ref{yield1} we only 
show results for the 39 nuclear species that had stellar yields 
with mean magnitudes greater than $\sim 10^{-10}$ and hence neglect 
many short-lived unstable isotopes that are included in the 
nucleosynthesis network as well as in the tables that are 
available for download.   Using this criterion we have included 
results for the important radio-nuclides \iso{26}Al and 
\iso{60}Fe, noting that we set the production factors to be
zero for these models owing to an initial zero abundance. 

\begin{figure}[t]
\begin{center}
\includegraphics[scale=0.5, angle=90]{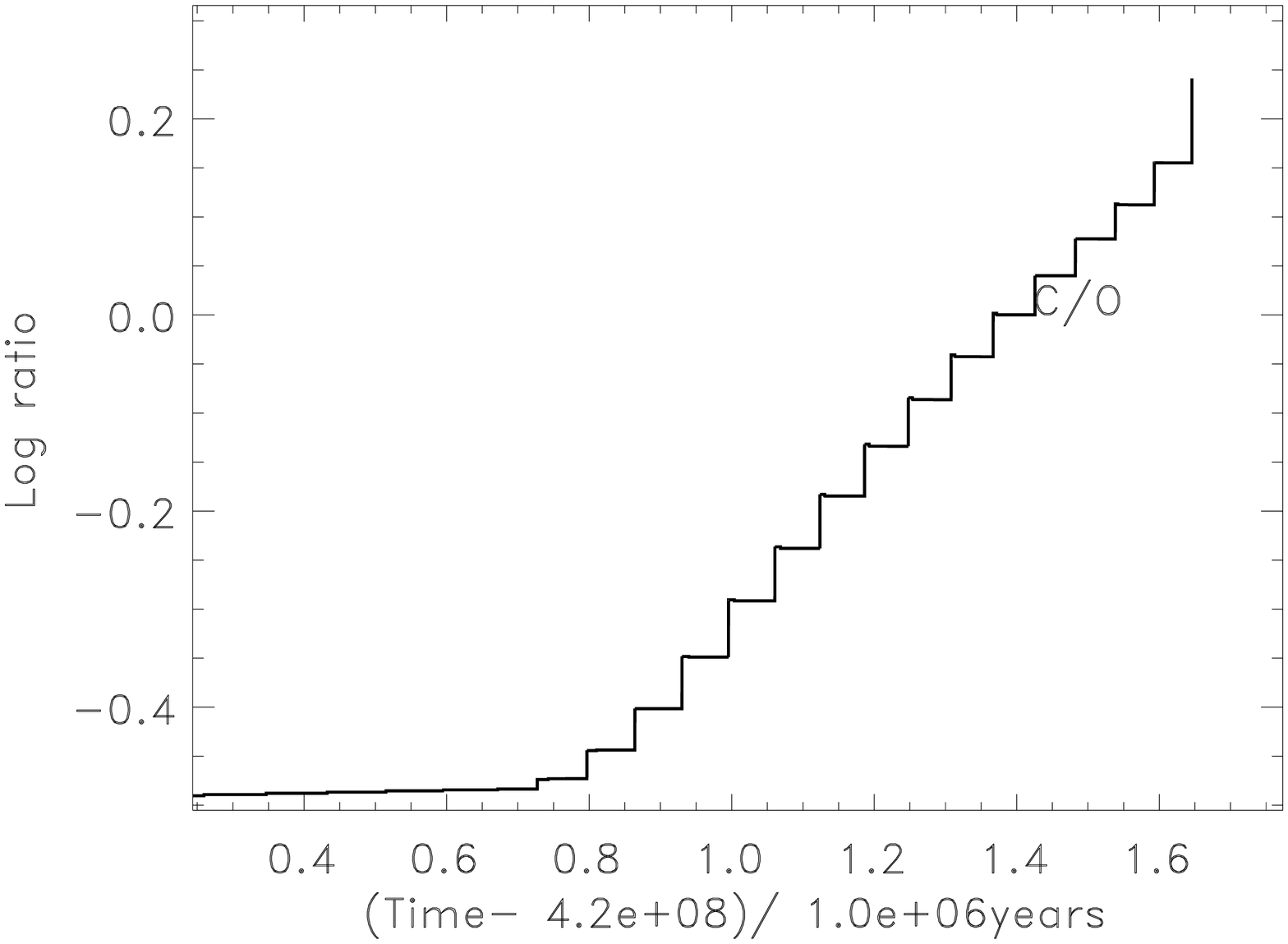}
\caption{The evolution of the C/O ratio at the surface during the TP-AGB phase
   of the 3$M_{\odot}$, $Z = 0.02$. The final C/O ratio for this model is 1.43.
}\label{m3z02CO}
\end{center}
\end{figure}

For the 3$\Msun$, $Z = 0.02$ model we estimate that the last TP will occur
when the total stellar mass was $\sim 1.35\Msun$ and the envelope 
$\sim 0.66\Msun$.
From inspection of Table~\ref{example2} it is clear that this last TDU
episode occurs at a much smaller envelope mass than all previous mixing
episodes, where $M_{\rm env} \gtrsim 2\Msun$. Consequently the change
to the surface abundances was larger and this can be seen in 
Fig.~\ref{m3z02CO} where we show the evolution of the C/O ratio at 
the surface. To see what effect this last TDU episode has on the 
stellar yields we compare the nett yield of \iso{12}C. Without this 
final TDU episode the nett yield of \iso{12}C is
1.354$\times 10^{-2}\Msun$ whereas with that final mixing event the 
yield is 1.496$\times 10^{-2}\Msun$, an increase of about 11\%.
On the other hand the change to the final surface abundance of \iso{12}C
is about 20\%.  

Before we move on it is important to discuss the effect of the
initial mass on the degree of chemical enrichment.
It was mentioned above that the 6$\Msun$
$Z=0.02$ model had an extra 5 TPs left, given the final envelope mass 
of 1.53$\Msun$ from the end of the detailed calculation. 
While it would seem obvious that a greater number of remaining TPs 
would lead to
a larger enrichment of the envelope, we need to remember that the
He-intershell is about a factor of 10 smaller in an intermediate-mass
AGB star. For this reason the amount of matter dredged into the envelope
at each TDU episode is smaller, regardless of the value of $\lambda$.
This can be seem from inspection of the tables given in \S\ref{modelresults}
where the maximum mass of the pulse-driven pocket (approximately 
equal to the mass of the intershell at the TP) is 0.016$\Msun$ 
for the 3$\Msun$, $Z=0.02$ model whereas this value is reduced to 
0.0033$\Msun$ in the 6$\Msun$, $Z=0.02$ case. The mass of 
the envelope in the 6$\Msun$ model was also larger at each TDU
episode resulting in more dilution of dredged-up material.
For this massive AGB model, the final surface abundance of 
\iso{12}C increased by about 26\% as a consequence of the 5
extra TPs, only marginally more than the 20\% increase from 
only {\bf one} TP for the 3$\Msun$, $Z=0.02$ case, 
as discussed above.

The stellar yields that we present here may be slightly different 
to those presented in \citet{karakasThesis} because we have improved
the synthetic AGB algorithm used to estimate the surface enrichment 
from the final TPs. The main improvement is taking intershell 
compositions for the last TP for all 74 species from the detailed 
stellar models. Previously we estimated the intershell 
compositions for about 15 species (e.g. \iso{4}He, \iso{12}C, \iso{16}O,
\iso{22}Ne etc) and assumed the rest did not change. We also did not
have intershell compositions for each mass and $Z$, so in some cases
we assumed that compositions appropriate for e.g. the 3$\Msun$, $Z=0.004$
model were appropriate for the 2.5 and 3.5$\Msun$ models of the 
same $Z$.  We have also improved the input data required for the 
synthetic code, e.g.  the final computed interpulse period
and TDU efficiency. These improvements
change the final remnant mass (at least slightly) and this also 
results in slight changes to the stellar yields.

\subsection{Planetary nebulae}

After the outer envelope has been ejected the stars evolve 
through the brief post-AGB and planetary nebula (PN) phases of the 
Hertzsprung-Russell diagram \citep{vanwinckel03}. The glowing
gas of the nebula is the remnant of the outer convective envelope 
of the progenitor AGB star and hence its composition should reflect 
the nucleosynthesis that took place during the final few 
TPs on the TP-AGB. 

Planetary nebulae are important because their compositions can be
determined fairly accurately (at least for some species); 
because they are bright, abundances can be derived from PNe 
as far away as M31, and because the gas is ionized by UV photons 
we obtain information about elements normally very difficult
to study, in particular the noble gases He, Ne, Ar, as well as C, N, O 
\citep{kaler78,henry89,dopita97,stanghellini00}. Abundances of 
heavy elements synthesized by the $s$ process including Ge, Se, Kr, Xe
Br and Ba \citep{sterling02,sterling07b,sharpee07} 
are also observed to be enhanced in some nebulae. Owing to the importance 
of PNe abundances we have included stellar yields for PNe in particular.
These yields are the abundance by number in the matter lost from 
the star from the final two TPs and as such are good for comparison
to the composition of the nebular gas for PNe in the 
Galaxy, LMC and SMC.  We have included PN yields
for the low-metallicity $Z= 0.0001$ models but few (if any) PN 
around today are expected to have evolved from such a low-metallicity
population of stars. PNe in the Milky Way Galaxy are 
predominantly found in the disk with oxygen abundances\footnote{used
as a proxy for metallicities because Fe is difficult to measure owing 
to condensation into dust.} in the range 
$7.73 \lesssim \log \epsilon ( \, {\rm O} \, ) \lesssim 9.09$ 
\citep{stanghellini06} compared to the solar value of 8.66
\citep{asplund05}.  Only a few PN have been found in the
halo and there are only a handful ($N \sim 12$) known that 
are defined to be metal-poor, with oxygen abundances less 
than a factor of 4 below solar \citep{dinerstein03}.

In Table~\ref{pnyield} we show an example of 
the yields available for a PN originating from a 3$\Msun$ 
progenitor of solar composition. As for the yields
presented previously we compute two sets for models with
synthetic TPs and for this case we show the PN set with the
final synthetic TP included. The first entry in each on-line
table is the amount of matter expelled during the last two 
TPs, here we include this value in the table caption.
The columns contain the species $i$, the atomic weight $A(i)$, 
followed by the nett yield, the mass lost, the initial 
mass fraction, and the average mass fraction 
integrated over the final two TPs.
The next quantity is the abundance (by number) of the matter 
lost in the wind over the final two TPs, compared to the number of 
hydrogen atoms, $N_{i}/N_{\rm H}$. This quantity is 
often used in the literature in the form 
$\log \epsilon =  \log (N_{i}/N_{\rm H}) + 12$.   

Comparing results for \iso{12}C from Table~\ref{pnyield} to
Table~\ref{yield1} we see that the nett yield is lower 
in the PN table by $\sim 27$\%, this is consequence of 
less matter being lost in the final two pulses than over 
the whole lifetime, even though the average \iso{12}C 
abundance is higher by $\sim 8$\%. Remember that our 
definition  of a stellar yield defined in 
Equation~\ref{eq:netyield} essentially subtracts the 
amount of $i$ in the wind lost from the star (i.e. 
$\approx M_{\rm lost}^{\rm total} \times \langle X(i) \rangle$), 
by the amount of $i$ that would have been initially 
present in that wind (i.e. 
$M_{\rm lost}^{\rm total} \times X_{0}(i)$). Note
that the 3$\Msun$, $Z = 0.02$ model lost $1.508\Msun$ 
over the last two TPs compared to
$3.0 - M_{\rm f} = 2.318\Msun$ over the entire lifetime 
(see Table~\ref{missedTPs} for the final mass).

That the yield difference was not larger is a consequence
of the VW93 mass-loss prescription. This prescription
ensures that the mass-loss rate is low until the superwind
begins at which stage most of the envelope is lost in
a few TPs. For the 5$\Msun$, $Z= 0.02$ model we computed
an evolutionary sequence with the R75 mass-loss law
with $\eta = 3.5$. For this model the nett PN yield 
of \iso{12}C is 3.42 times lower than the total yield,
compared to $1.68$ times lower for the VW93 case.
The total amount of matter expelled is 0.75$\Msun$ and 
0.58$\Msun$, for the VW93 and R75 cases, respectively.
Note that the average \iso{12}C abundance in the 
wind expelled over the last two TPs increased  
in both models, where the R75 model saw a 66\% increase
compared to 63\% for the VW93 case.

\subsection{Limitations of the current models}

We discussed in the previous section that we did not include a partial
mixing zone at the deepest extent of each TDU episode to produce 
a \iso{13}C pocket. This can be considered a limitation 
since the neutrons produced by the \iso{13}C($\alpha,n$)\iso{16}O 
reaction can affect the composition
of the He-shell. However the light elements (hydrogen through to
phosphorous) are generally not significantly altered by neutron
captures, at least at solar metallicity.
There are exceptions to this, including the nucleosynthesis of \iso{19}F
that requires the operation of the \iso{13}C neutron 
source in low-mass AGB stars of $\sim 3\Msun$ \citep{jorissen92,lugaro04}. 
Sodium, the magnesium and silicon isotopes, and \iso{31}P are affected
by neutron captures during the TP where neutrons are released by the
\iso{22}Ne source at higher neutron densities ($\sim 10^{10}$ cm$^{-3}$), 
noting we include this in our computations. Neutron captures will impact 
more significantly on light-element nucleosynthesis in more massive or 
lower metallicity models, as pointed out by \citet{herwig04a,herwig04b}.

One limitation of our models, at least in terms of their use for galactic
chemical evolution studies,  are the initial C, N and O abundances
used for the $Z=0.008$ and 0.004 models. These are not scaled solar but
reflect the composition of the LMC and SMC, respectively. 
Our models are therefore very useful for comparing to the composition of
AGB stars and PN in the Large and Small Magellanic Clouds, which was
their intended use, but one must be careful when using the C, N and O yields for
comparison to populations that have a similar [Fe/H] content of the Clouds
but a different chemical history. We have started to compute yields assuming
a scaled-solar composition for those metallicities and these will be 
available in a future publication.

Another limitation is that our lowest metallicity models are
$Z = 10^{-4}$ whereas there are many new discoveries of low-mass
halo stars with [Fe/H] $\lesssim -3$ \citep{beers05}, some of
which could have been polluted by a former AGB companion.
We do not intend to compute AGB models with such low metallicities 
in the future but we refer the reader to Simon Campbell's 
models \citep{campbellThesis}. Besides that 7$\Msun$, $Z=0.0001$
super-AGB star we do not compute models for any more of these objects.
Yields for these stars should be available
in the future (Doherty, Lattanzio \& Siess, in preparation), noting
that models of these stars are proving to be even more 
challenging than for CO-core AGB models \citep{siess06}.

We do not include the effect of extra-mixing processes into
our computations, and by this we mean mixing events beyond those
predicted to occur by standard models such as ours e.g. the first
and second dredge-up, the TDU and HBB. Observations suggest that
extra-mixing processes reduce the \iso{12}C/\iso{13}C ratio in
low-mass ($m \lesssim 2\Msun$) red giants stars below that 
predicted by standard models \citep{charbonnel94}. Observations 
of AGB stars and pre-solar grain analysis indicate that a 
similar type of extra-mixing occurs in low-mass AGB stars 
\citep{boothroyd95,abia97,busso99,nollett03}.
While the physical mechanism for the extra-mixing is 
unknown\footnote{and this is the main reason why we 
do not include it! Another reason being that it is essential 
to have standard models to compare with before testing new
physics.} it is commonly thought to be rotation 
\citep[e.g.][]{palacios06}, although we also refer to the
mixing recently discovered by \citep{eggleton06} and 
investigated further by \citet{eggleton07} and 
\citep{charbonnel07}.

\subsection{Comparison with other yields}

\begin{figure}
\begin{center}
\begin{tabular}{c}
\includegraphics[scale=0.4, angle=270]{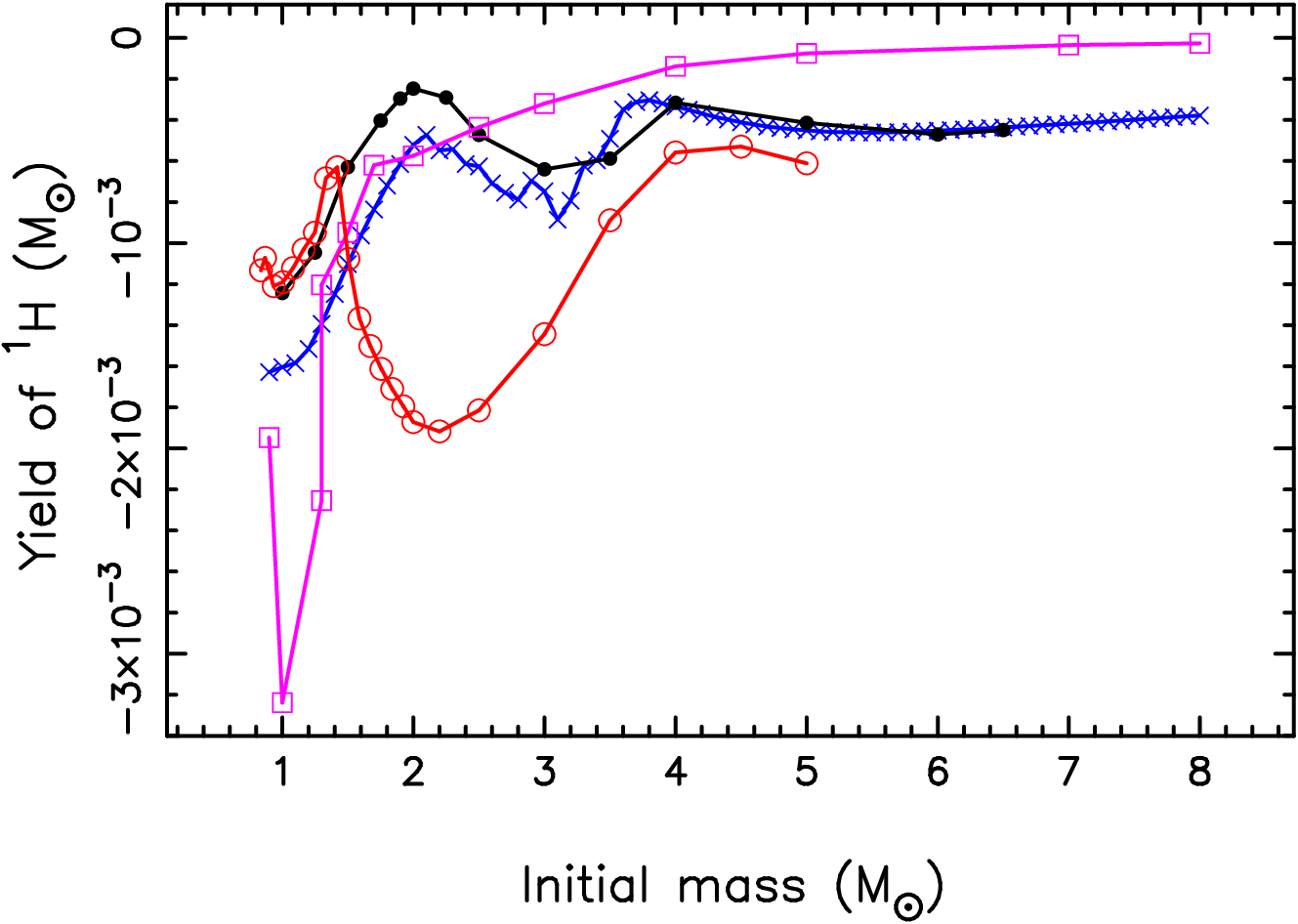}\\
\includegraphics[scale=0.4, angle=270]{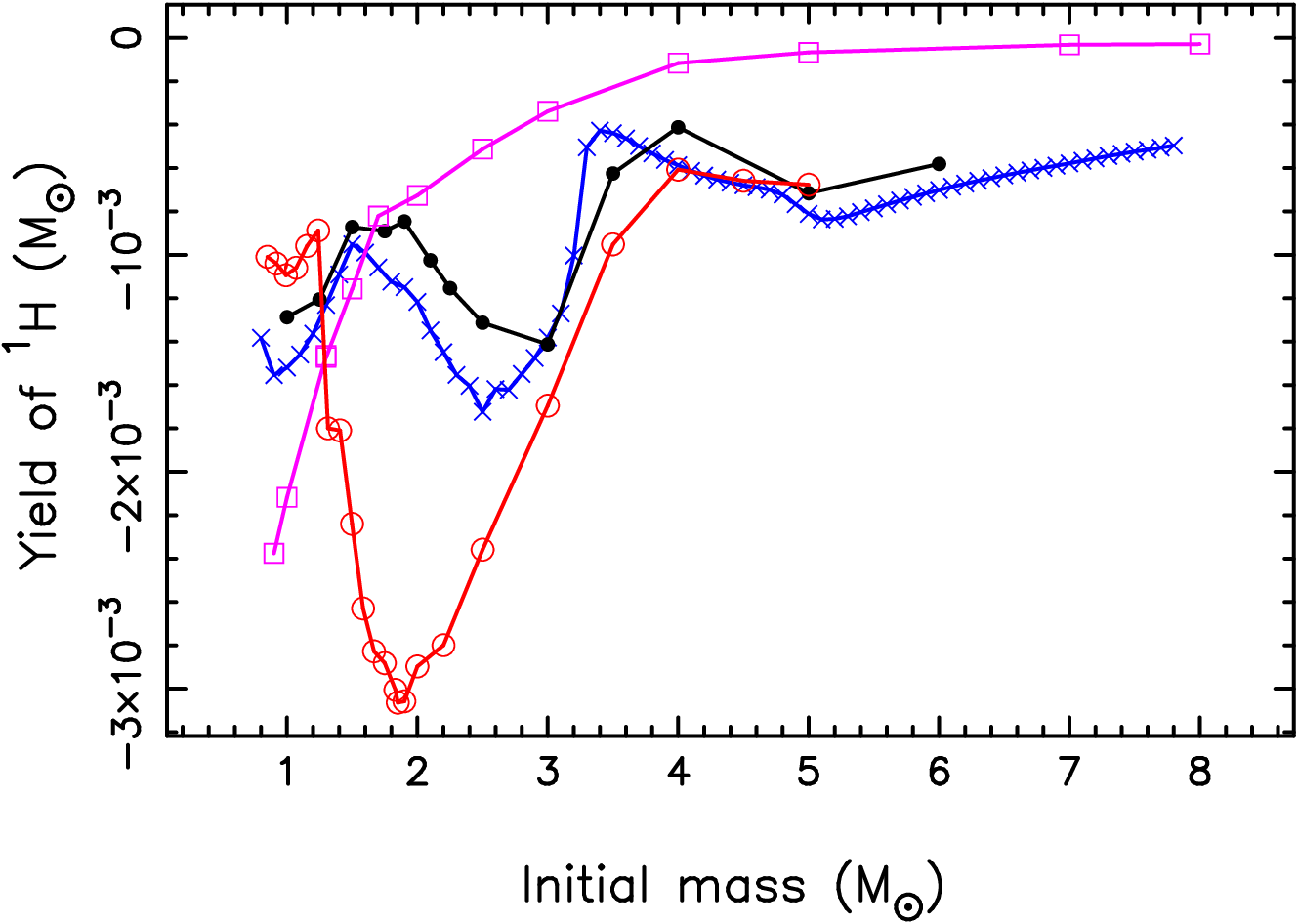}\\
\includegraphics[scale=0.4, angle=270]{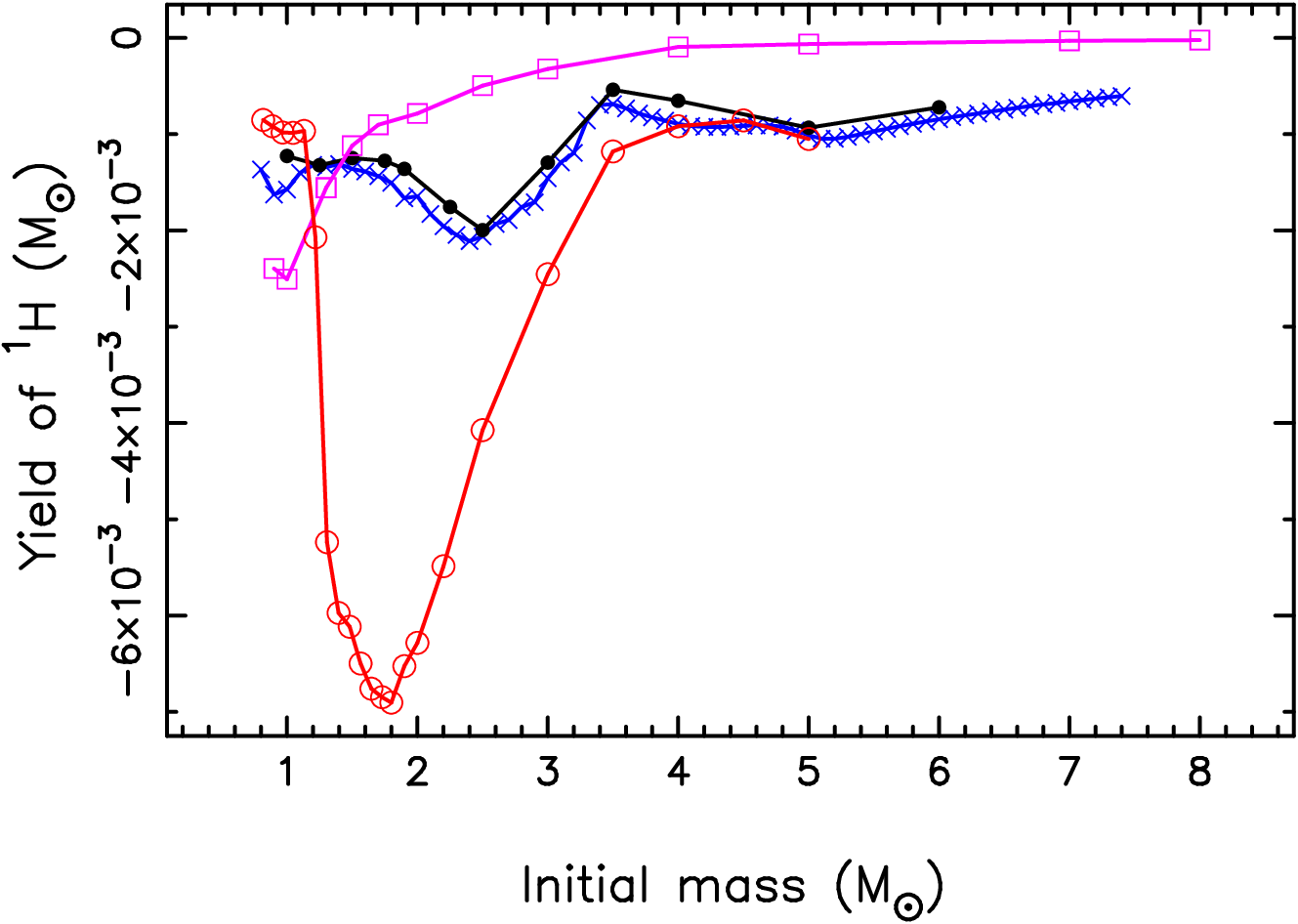}
\end{tabular}
\caption{Weighted yield of \iso{1}H as a function of the initial
mass for the $Z=0.02$ (top), the $Z=0.008$ (middle) and the 
$Z=0.004$ models (bottom). We show results from
our calculations (black solid points), \citet{vandenhoek97} (open
magenta squares), \citet{forestini97} (solid green squares), \citet{marigo01}
(open red circles), \citet{ventura02} (solid aqua triangles) and
\citet{izzard04b} (blue crosses). Forestini \& Charbonnel (1997)
and Ventura et al. (2002) do not provide yields for $Z=0.008$ and cover a
narrower mass range, between 3 and 6$\Msun$.
}\label{hyd}
\end{center}
\end{figure}

\begin{figure}
\begin{center}
\begin{tabular}{c}
\includegraphics[scale=0.4, angle=270]{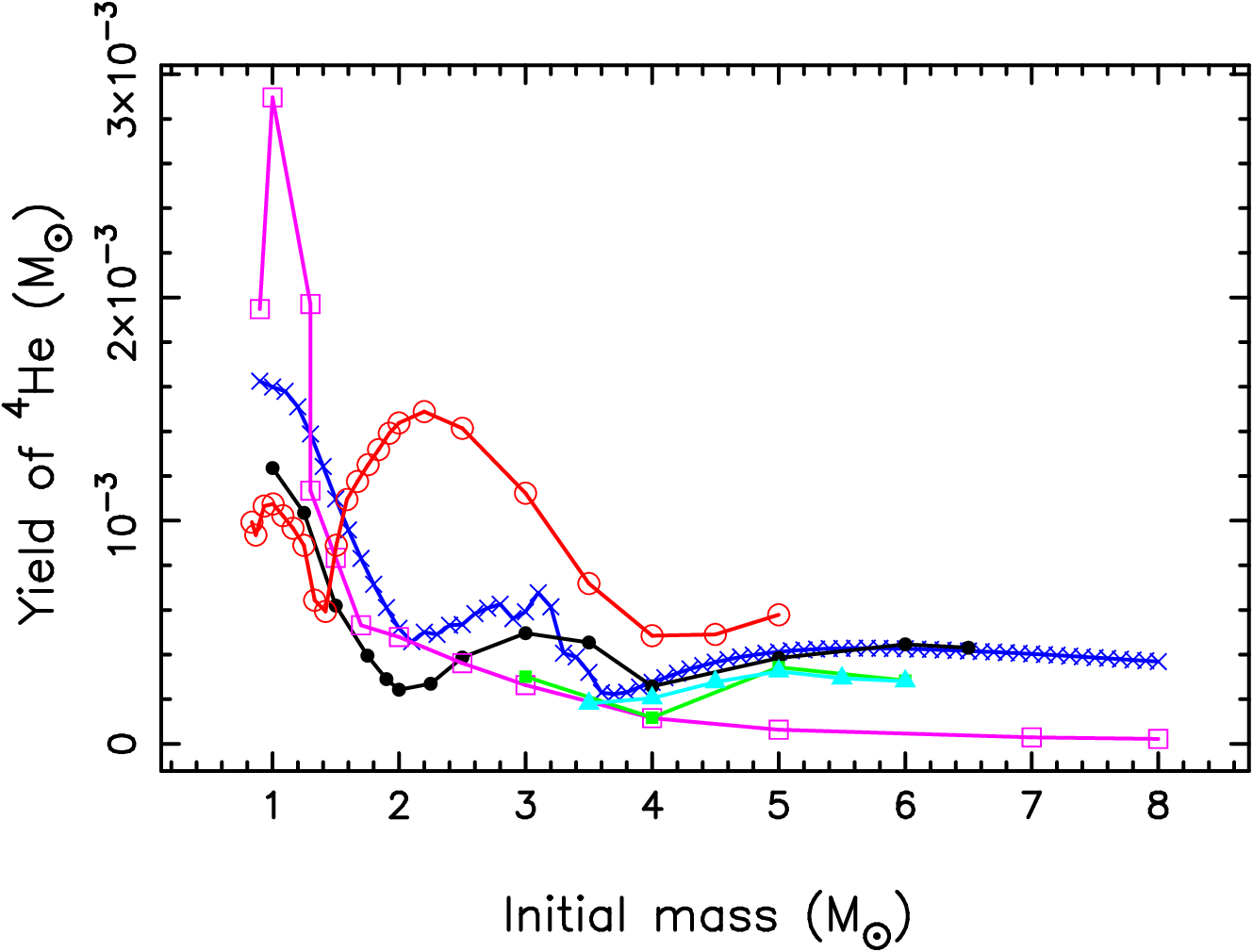}\\
\includegraphics[scale=0.4, angle=270]{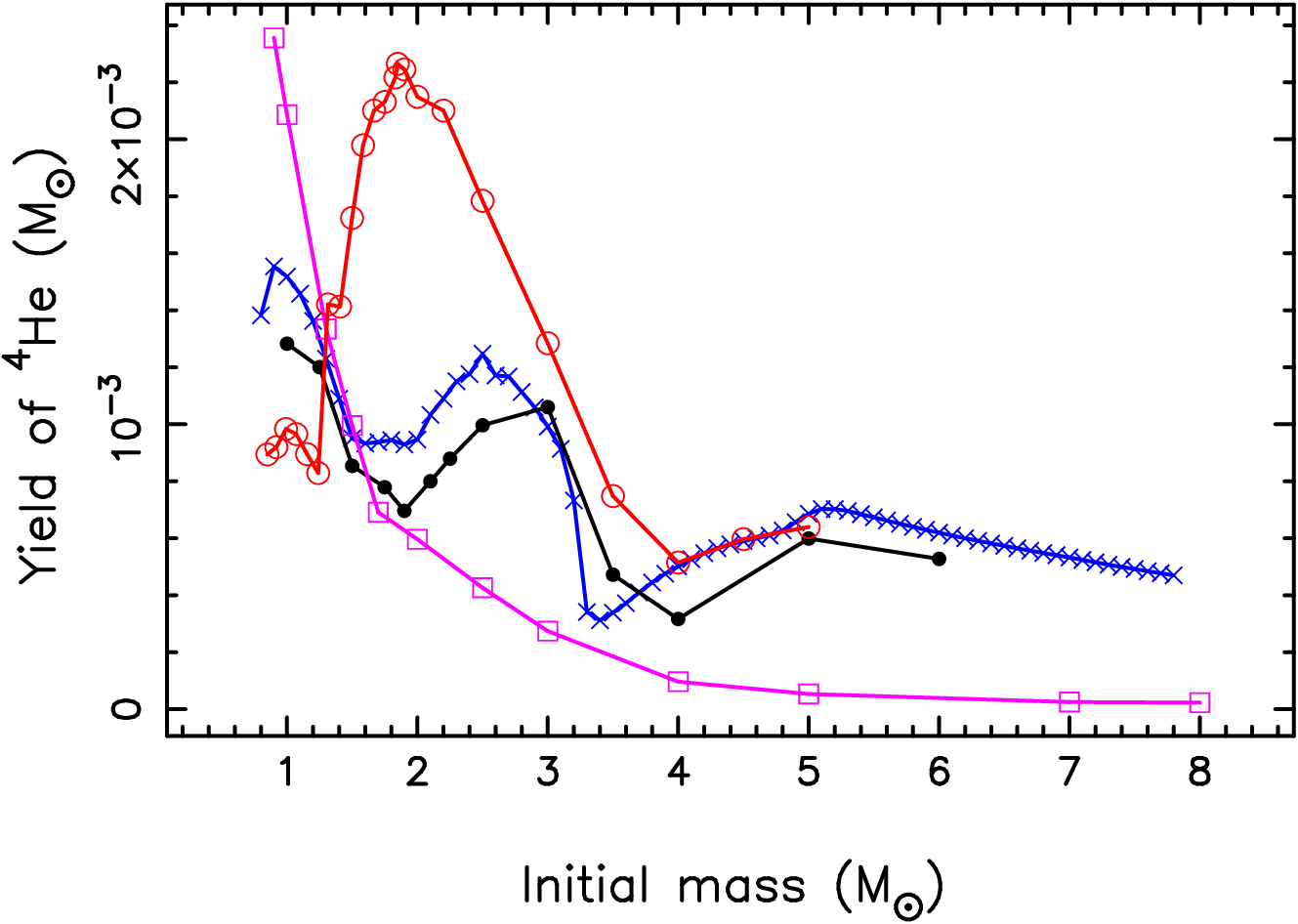}\\
\includegraphics[scale=0.4, angle=270]{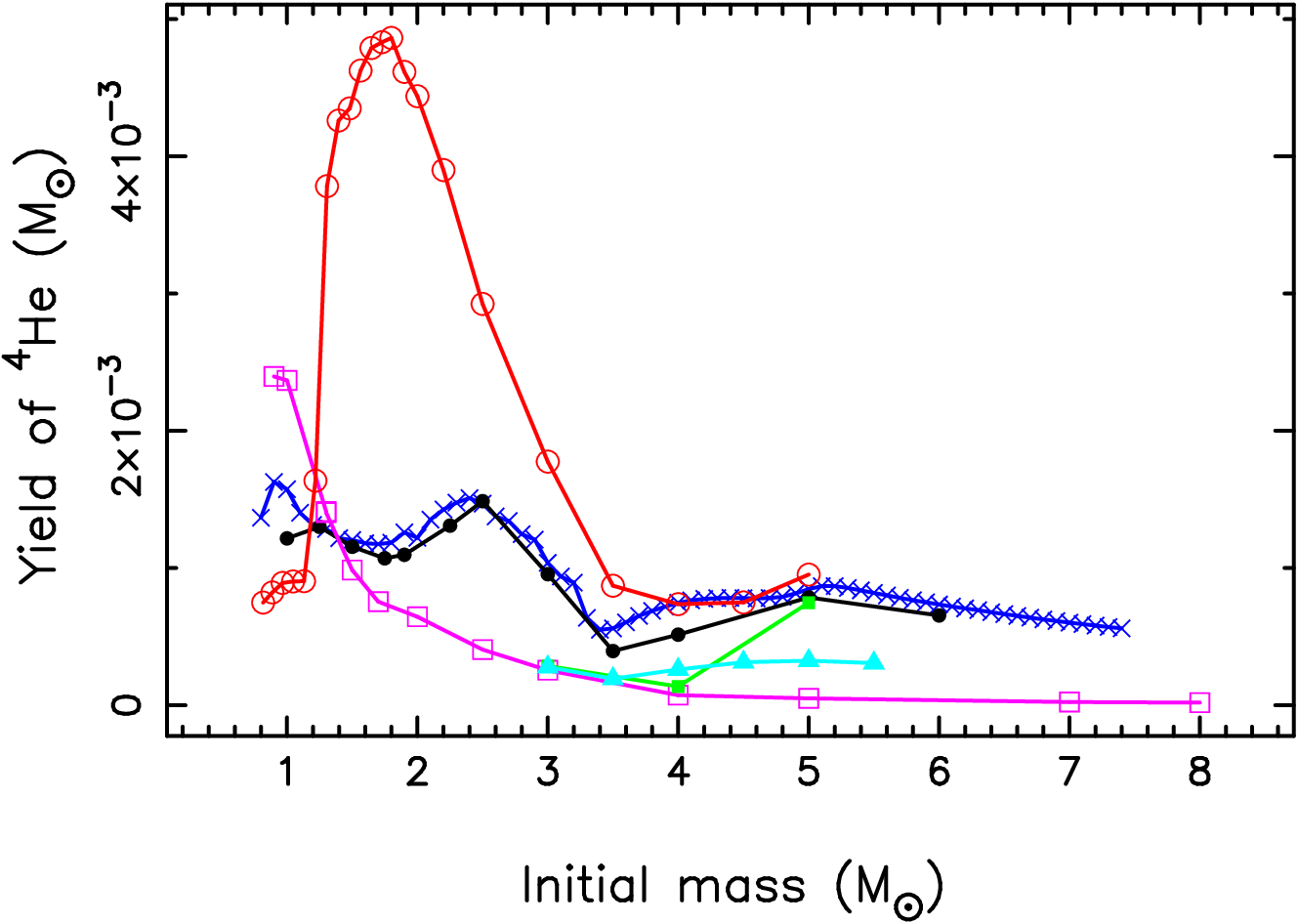}
\end{tabular}
\caption{Weighted yield of \iso{4}He as a function of the initial
mass for the $Z=0.02$ (top), the $Z=0.008$ (middle) and the 
$Z=0.004$ models (bottom). 
Symbols are the same as in Fig.~\ref{hyd}.
}\label{he4}
\end{center}
\end{figure}

\begin{figure}
\begin{center}
\begin{tabular}{c}
\includegraphics[scale=0.4, angle=270]{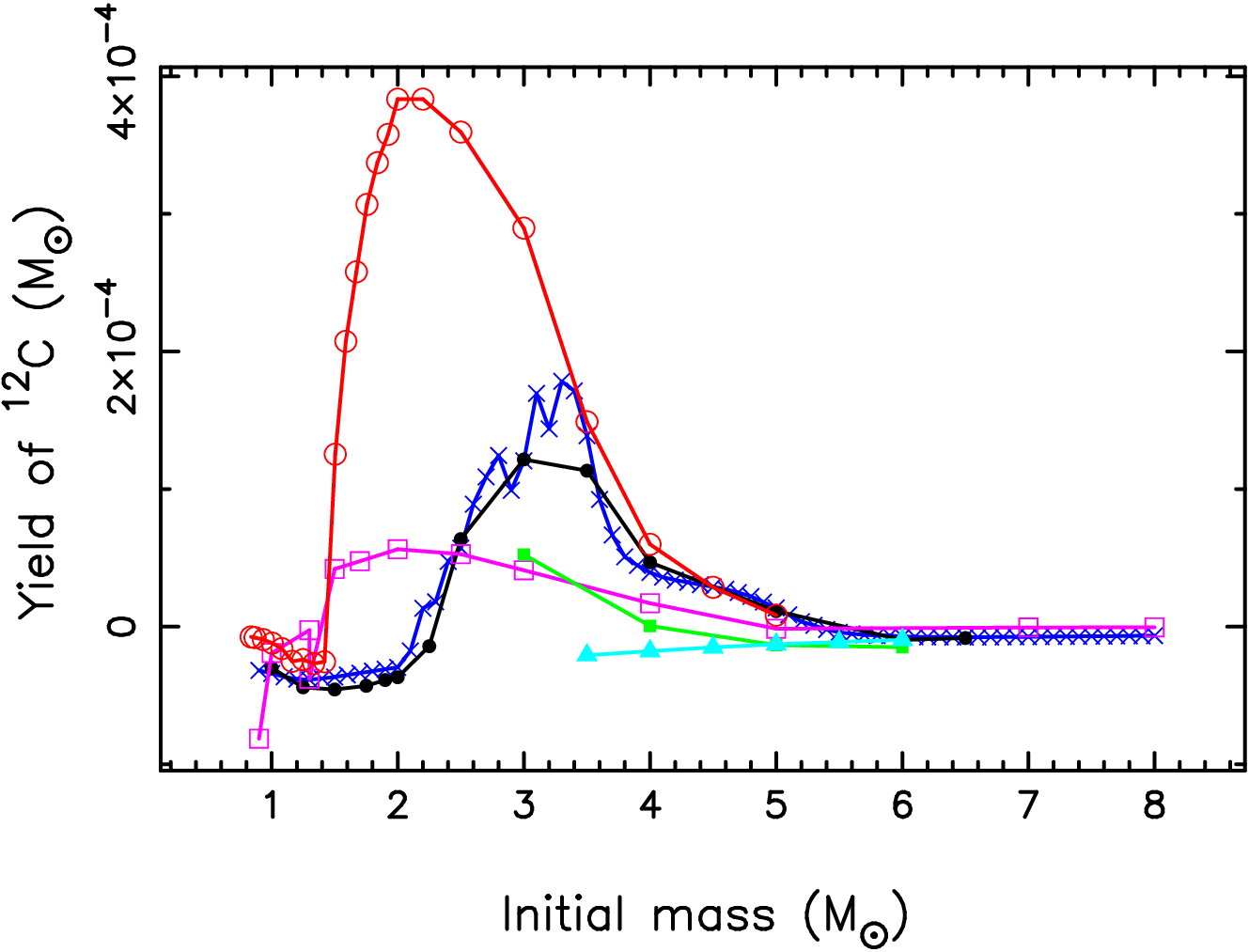}\\
\includegraphics[scale=0.4, angle=270]{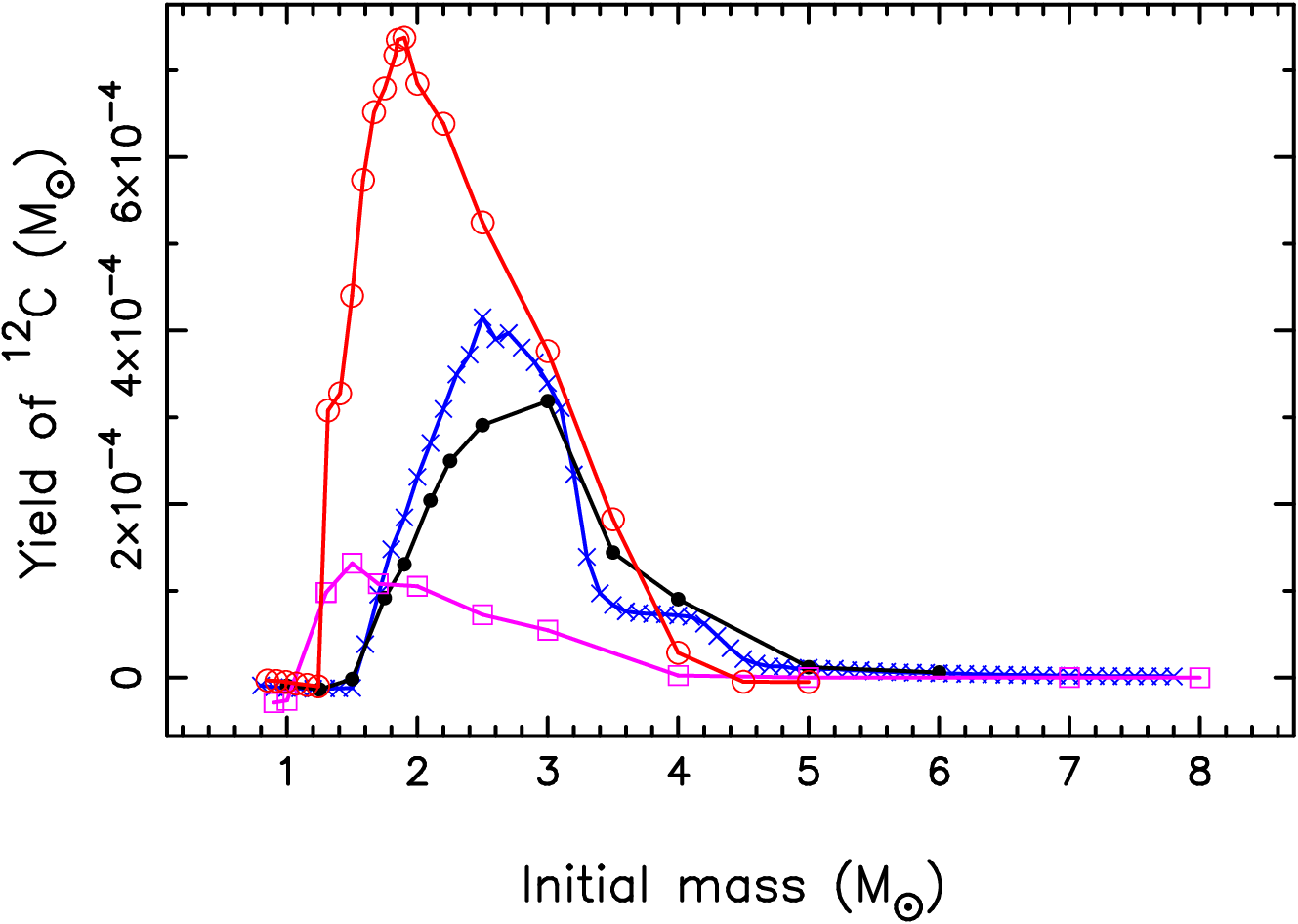}\\
\includegraphics[scale=0.4, angle=270]{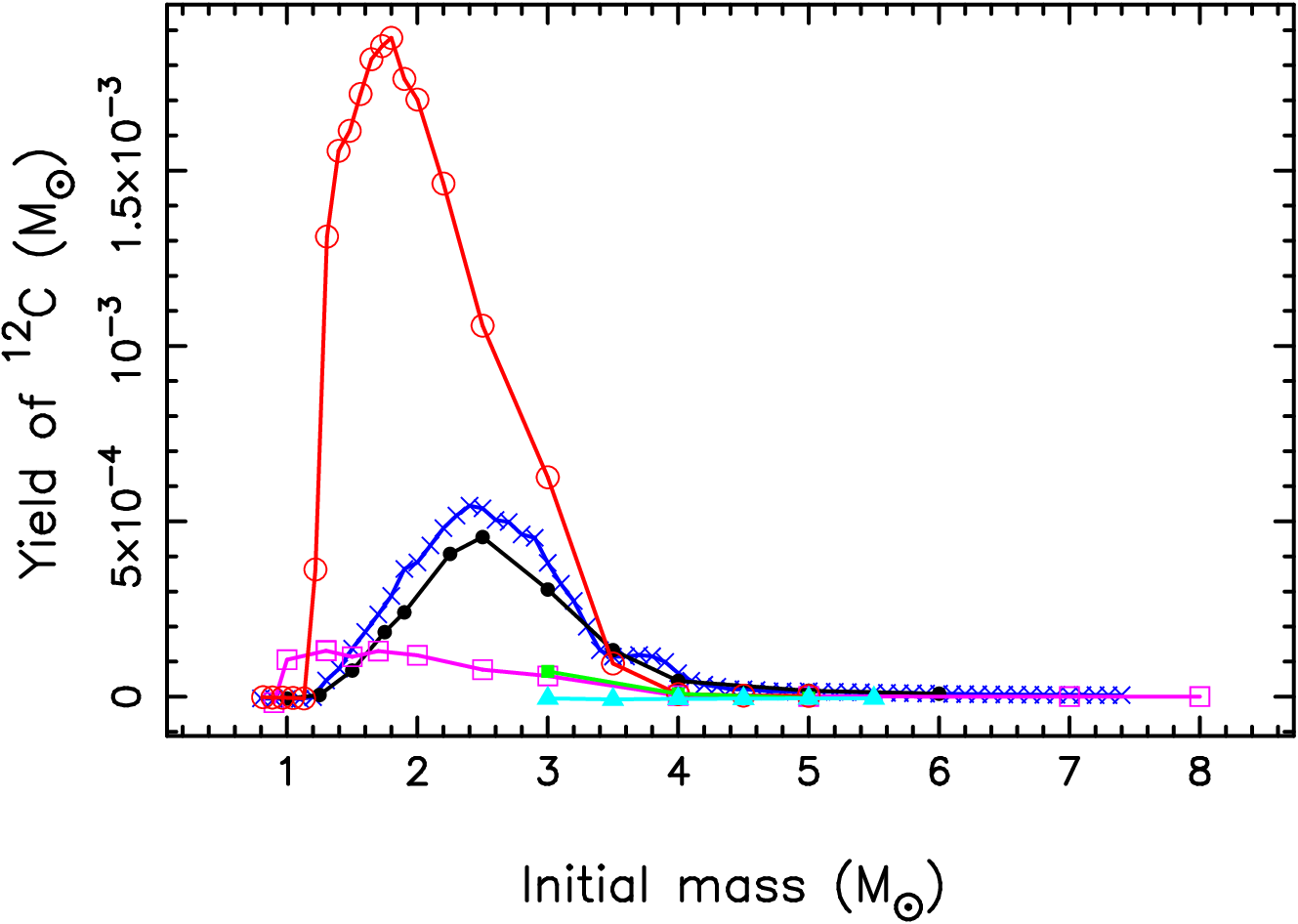}
\end{tabular}
\caption{Weighted yield of \iso{12}C as a function of the initial
mass for the $Z=0.02$ (top), the $Z=0.008$ (middle) and the 
$Z=0.004$ models (bottom). 
Symbols are the same as in Fig.~\ref{hyd}.
}\label{c12}
\end{center}
\end{figure}

\begin{figure} 
\begin{center}
\begin{tabular}{c}
\includegraphics[scale=0.4, angle=270]{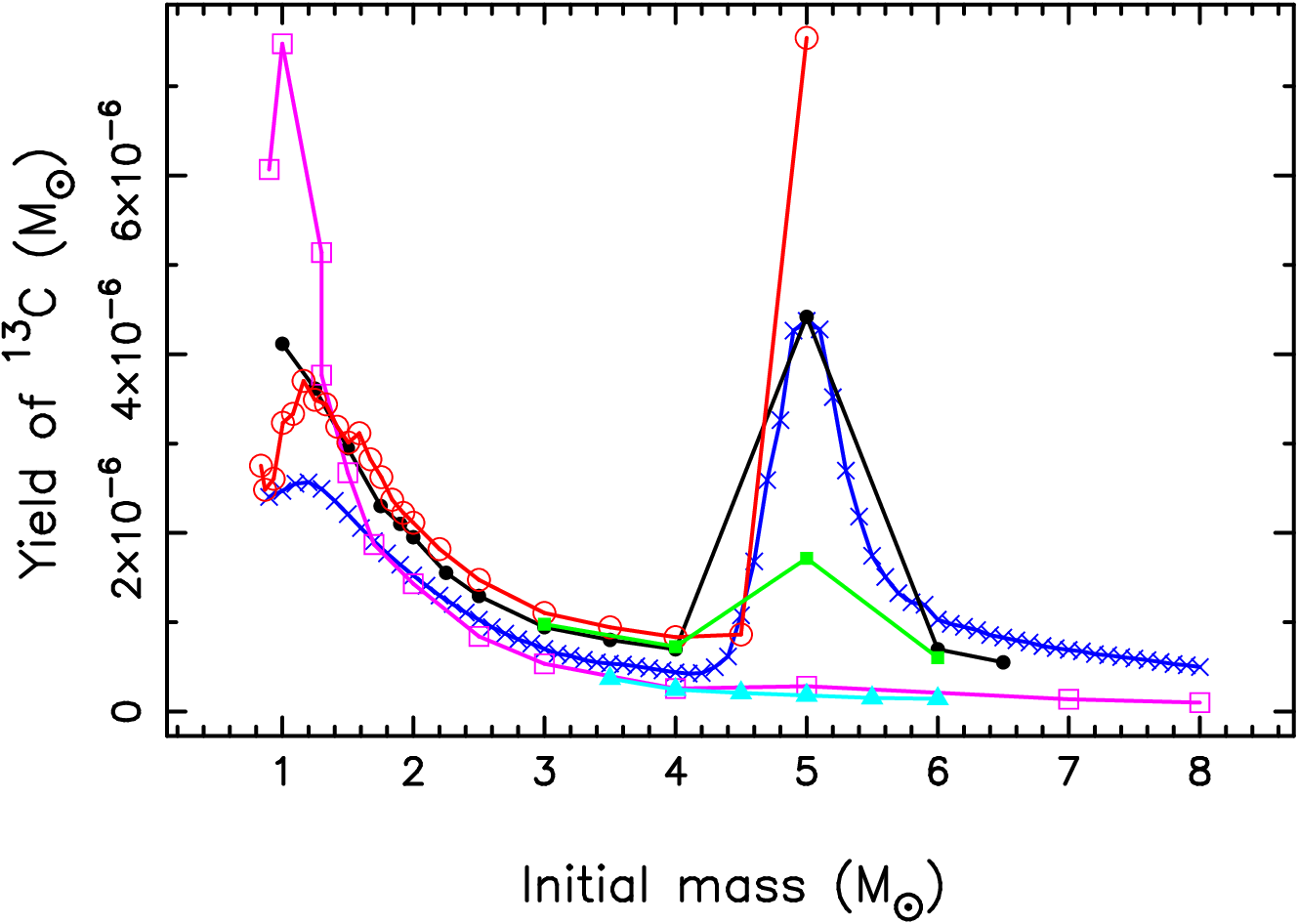}\\
\includegraphics[scale=0.4, angle=270]{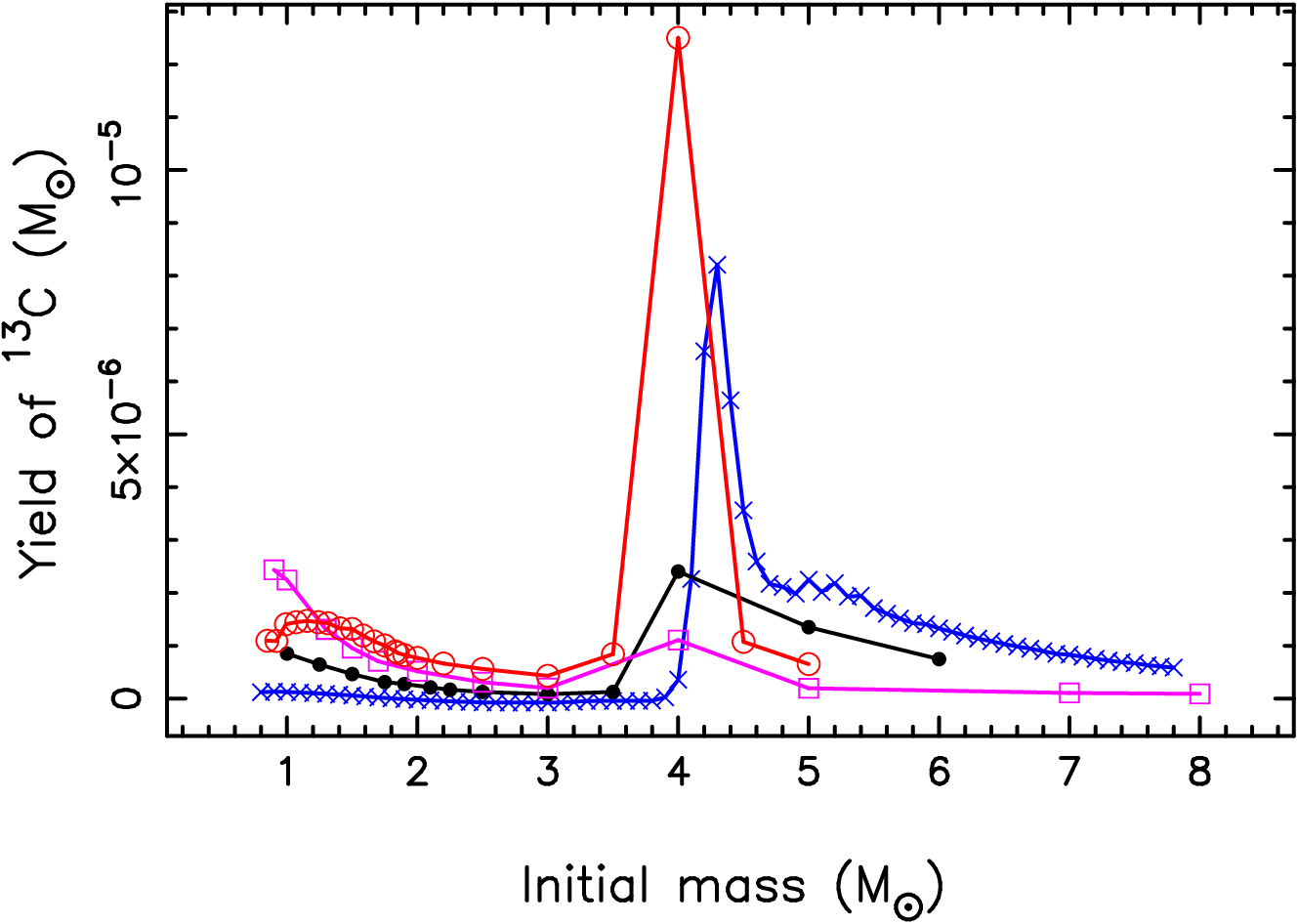}\\
\includegraphics[scale=0.4, angle=270]{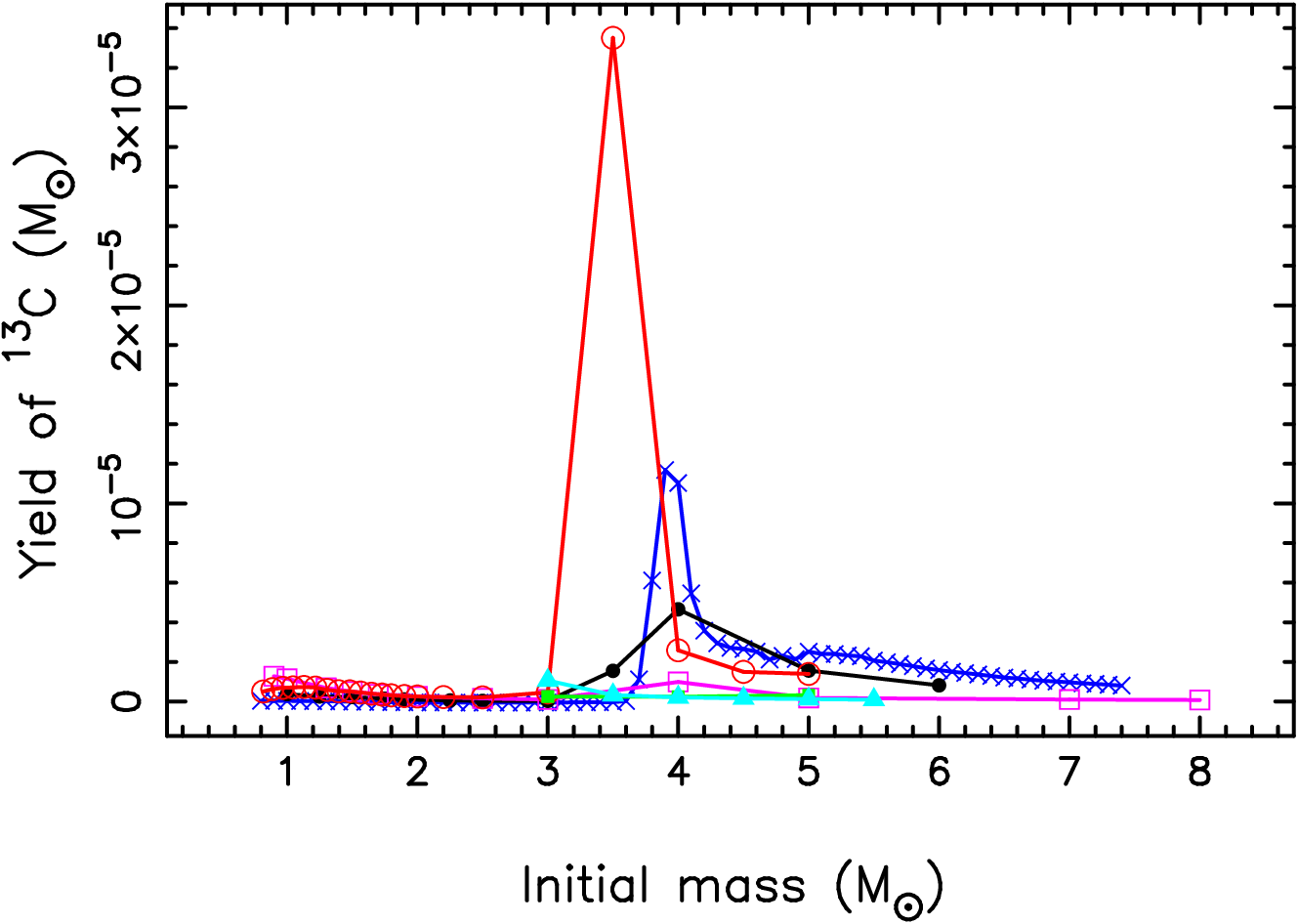}
\end{tabular}
\caption{Weighted yield of \iso{13}C as a function of the initial
mass for the $Z=0.02$ (top), the $Z=0.008$ (middle) and the 
$Z=0.004$ models (bottom). 
Symbols are the same as in Fig.~\ref{hyd}.
}\label{c13}
\end{center}
\end{figure}

\begin{figure} 
\begin{center}
\begin{tabular}{c}
\includegraphics[scale=0.4, angle=270]{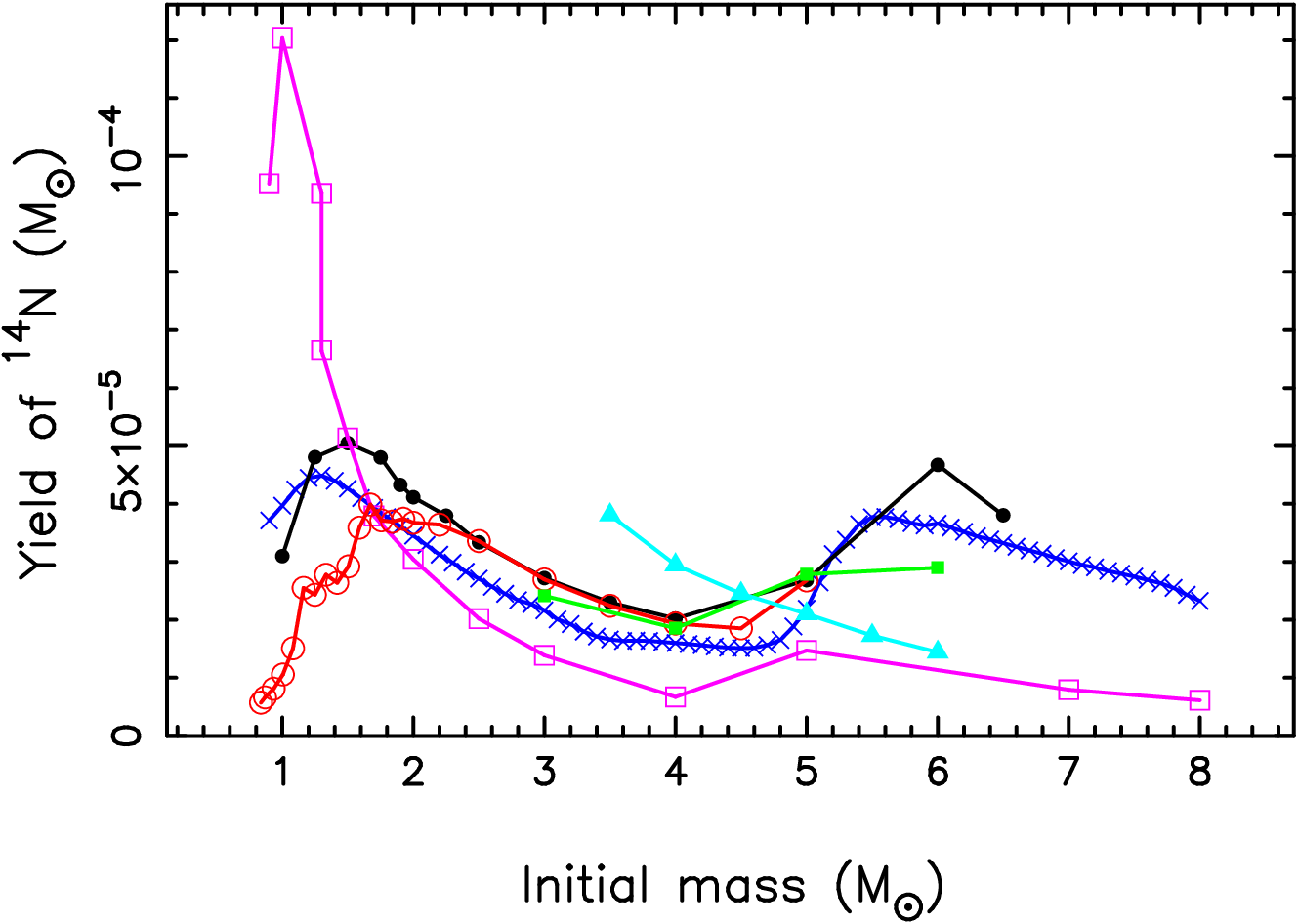}\\
\includegraphics[scale=0.4, angle=270]{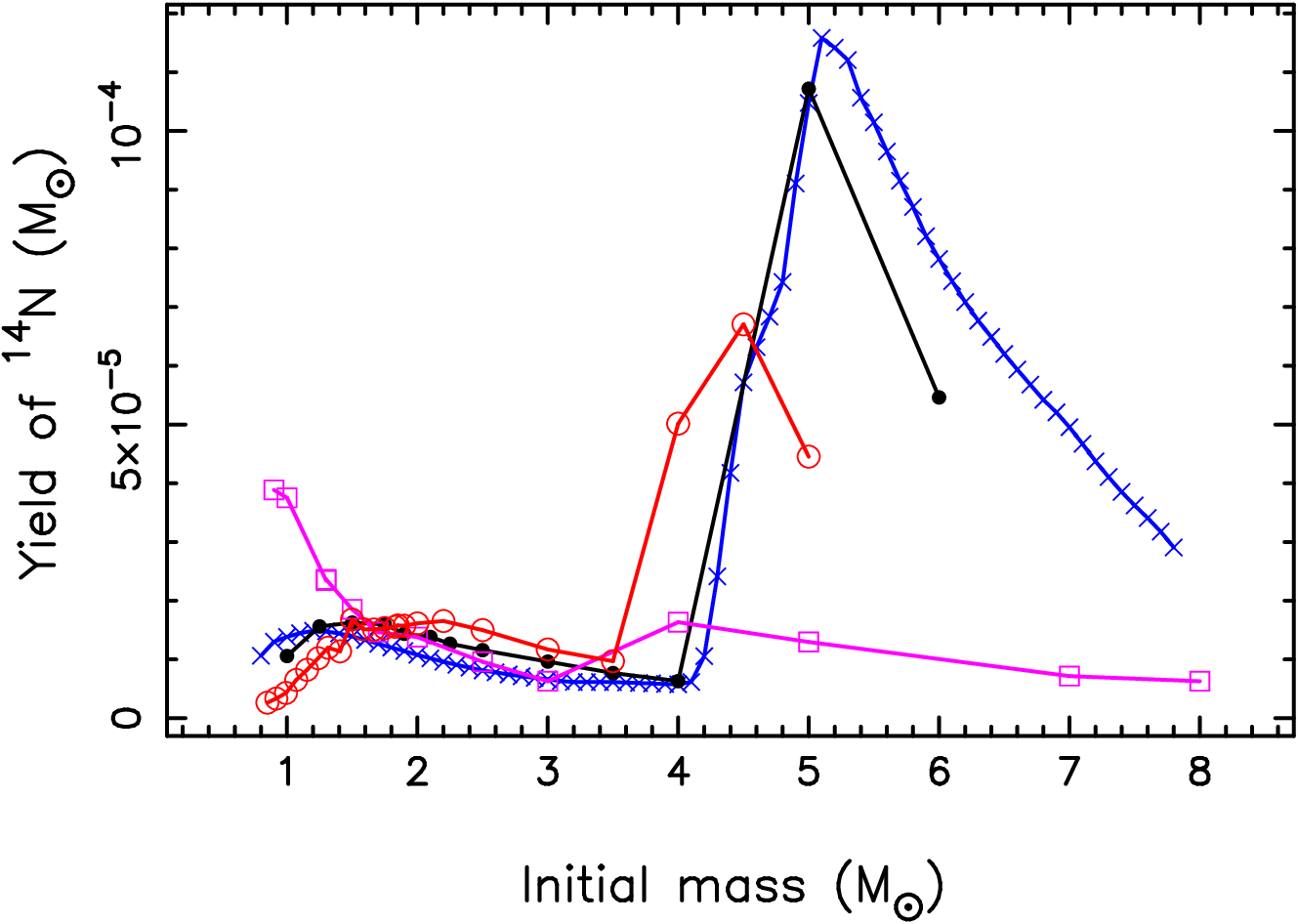}\\
\includegraphics[scale=0.4, angle=270]{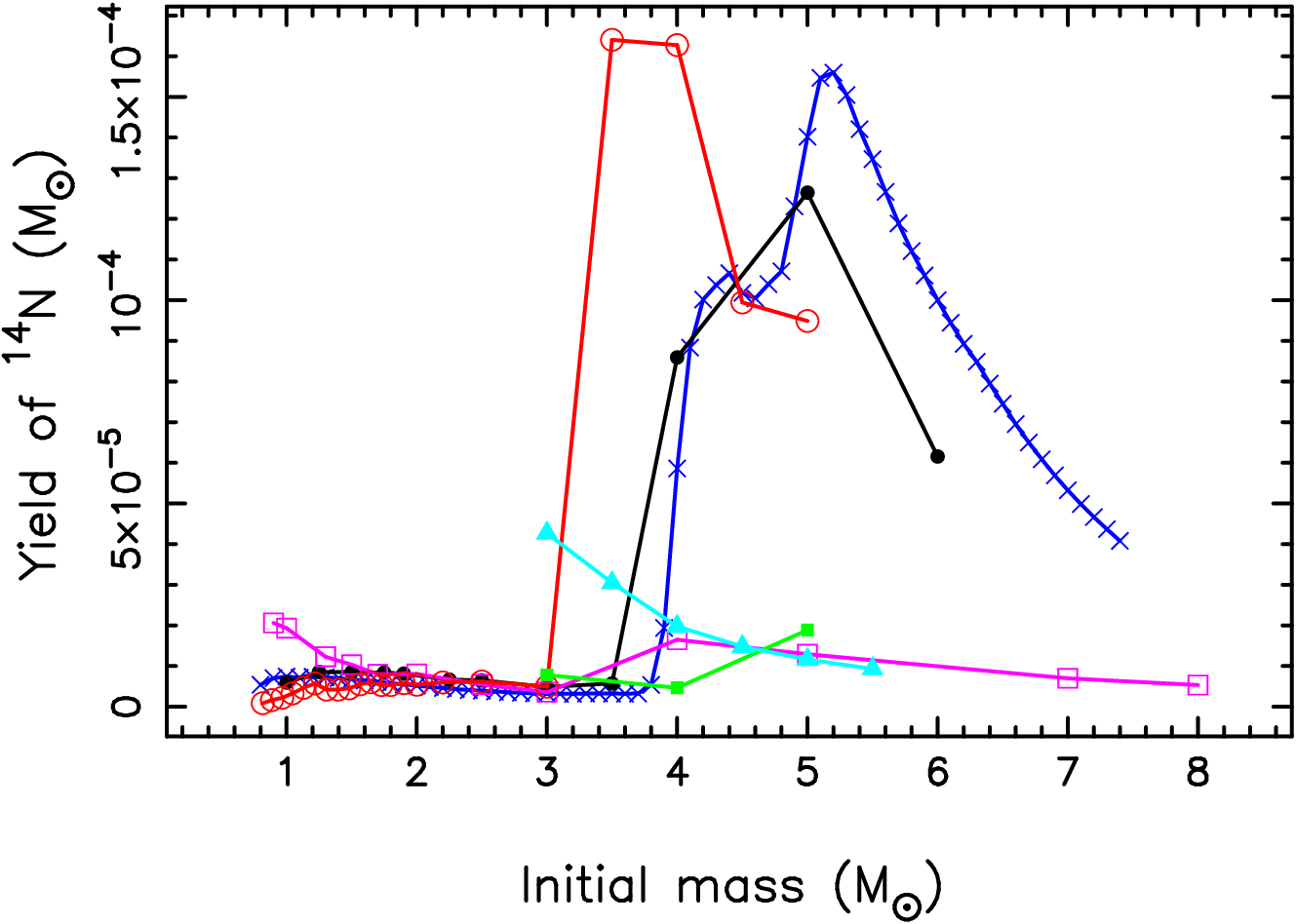}
\end{tabular}
\caption{Weighted yield of \iso{14}N as a function of the initial
mass for the $Z=0.02$ (top), the $Z=0.008$ (middle) and the 
$Z=0.004$ models (bottom). 
Symbols are the same as in Fig.~\ref{hyd}.
}\label{n14}
\end{center}
\end{figure}

\begin{figure} 
\begin{center}
\begin{tabular}{c}
\includegraphics[scale=0.4, angle=270]{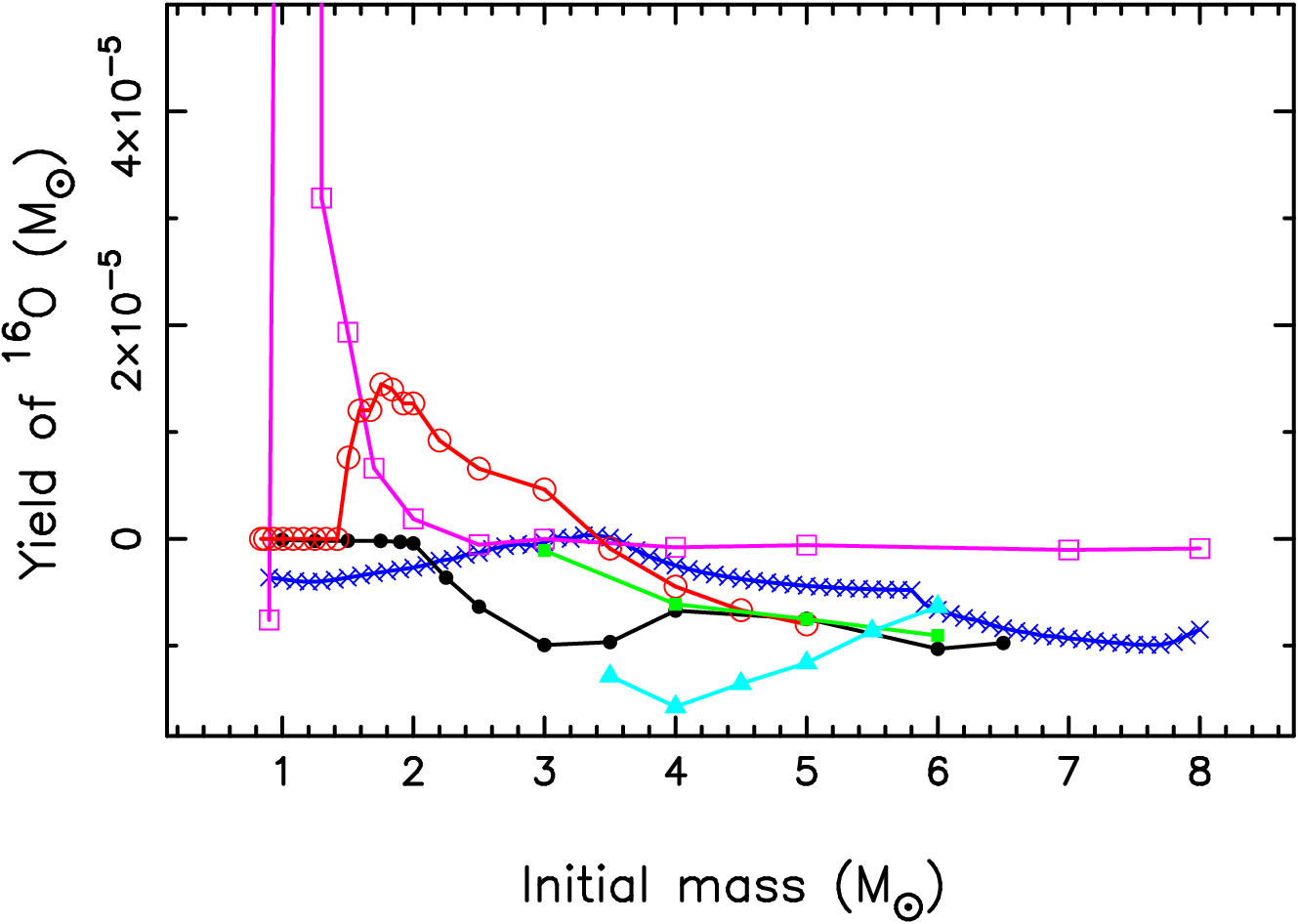}\\
\includegraphics[scale=0.4, angle=270]{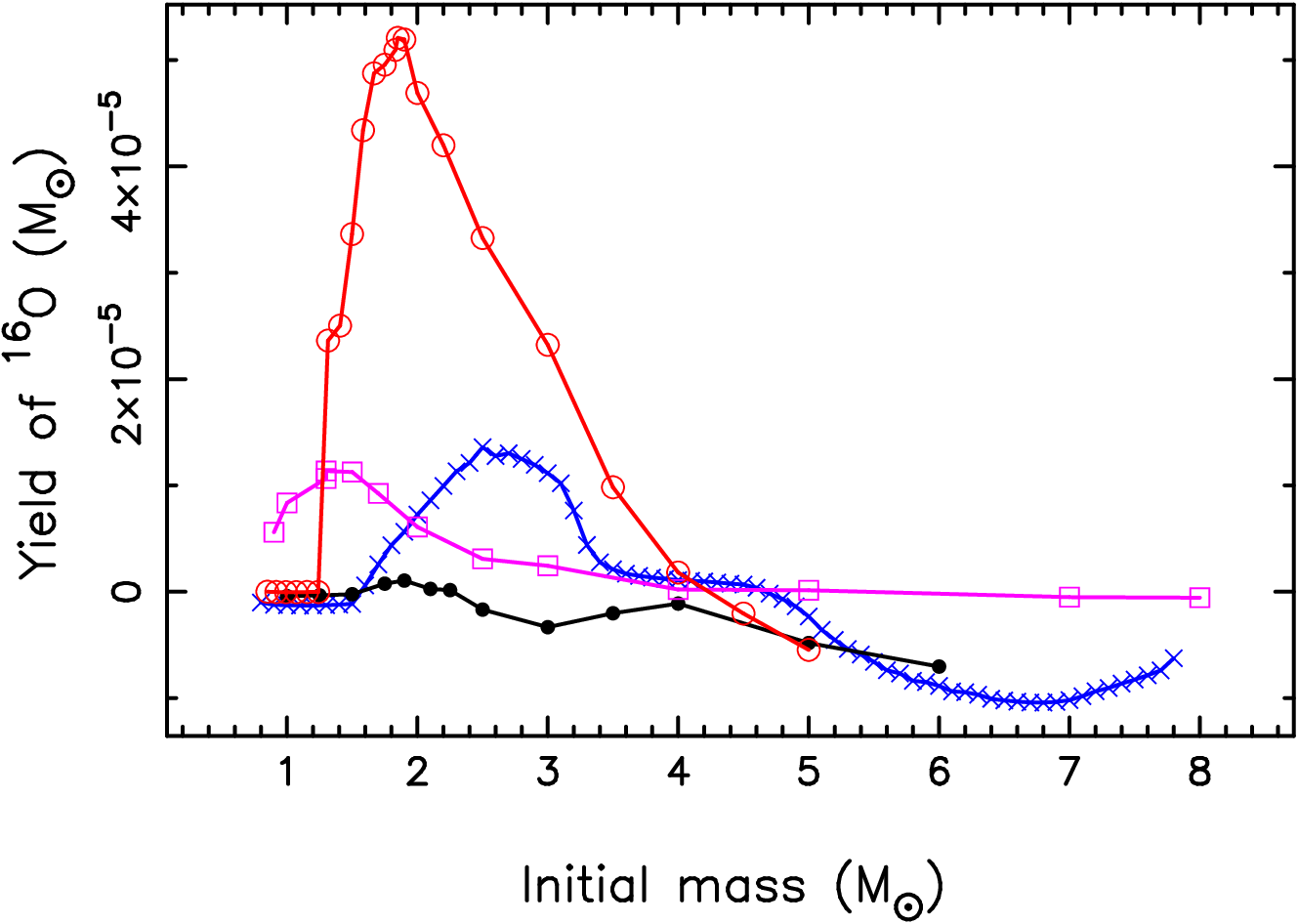}\\
\includegraphics[scale=0.4, angle=270]{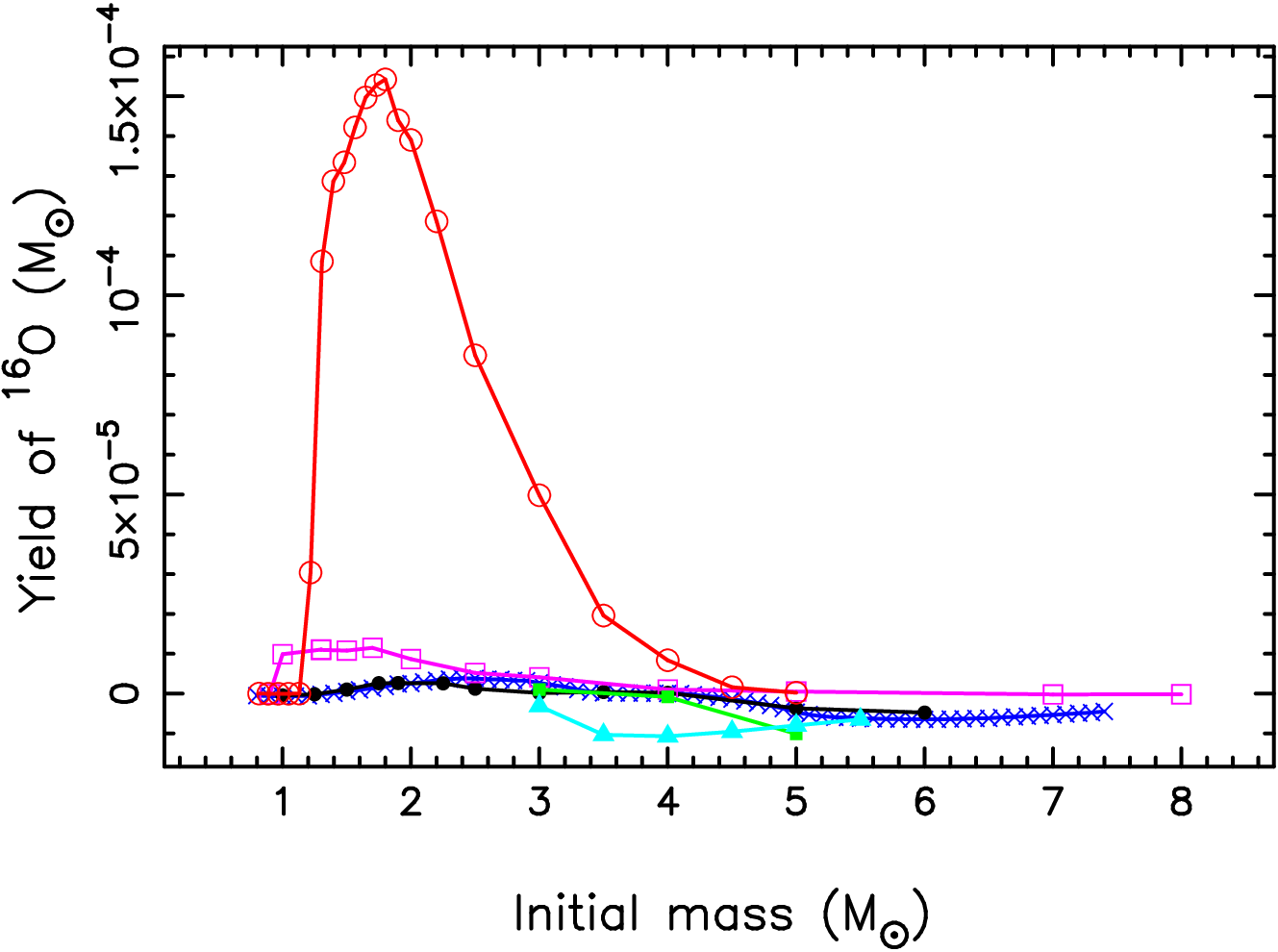}
\end{tabular}
\caption{Weighted yield of \iso{16}O as a function of the initial
mass for the $Z=0.02$ (top), the $Z=0.008$ (middle) and the 
$Z=0.004$ models (bottom). 
Symbols are the same as in Fig.~\ref{hyd}.
}\label{o16}
\end{center}
\end{figure}

\begin{figure}
\begin{center}
\begin{tabular}{c}
\includegraphics[scale=0.5, angle=270]{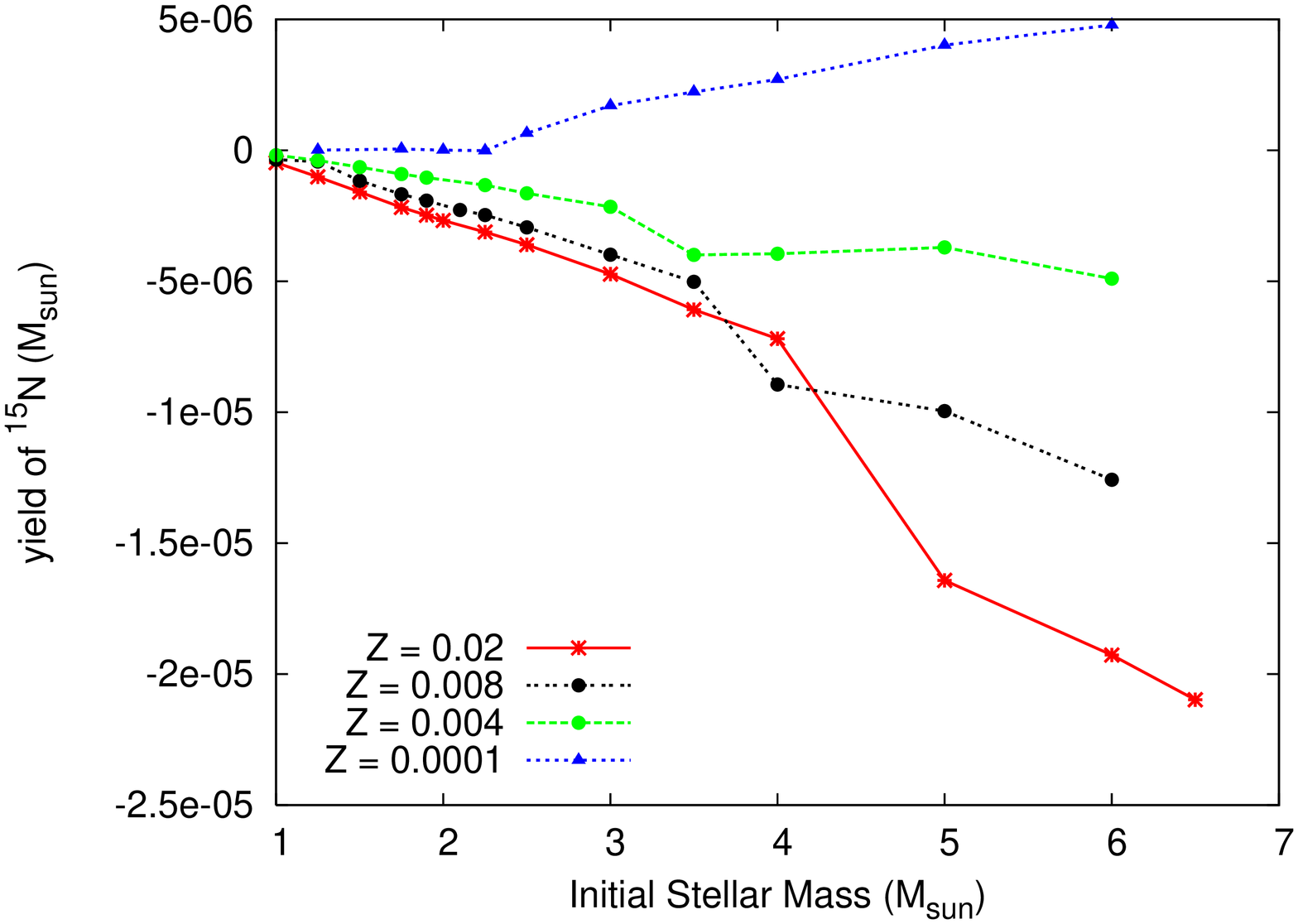} \\ 
\includegraphics[scale=0.5, angle=270]{f19-allz-bw.ps} 
\end{tabular}
\caption{Unweighted yield of \iso{15}N (top panel) and 
\iso{19}F (lower panel) as a function of the initial mass and 
metallicity $Z$.
}\label{f19}
\end{center}
\end{figure}

\begin{figure}
\begin{center}
\begin{tabular}{c}
\includegraphics[scale=0.5, angle=270]{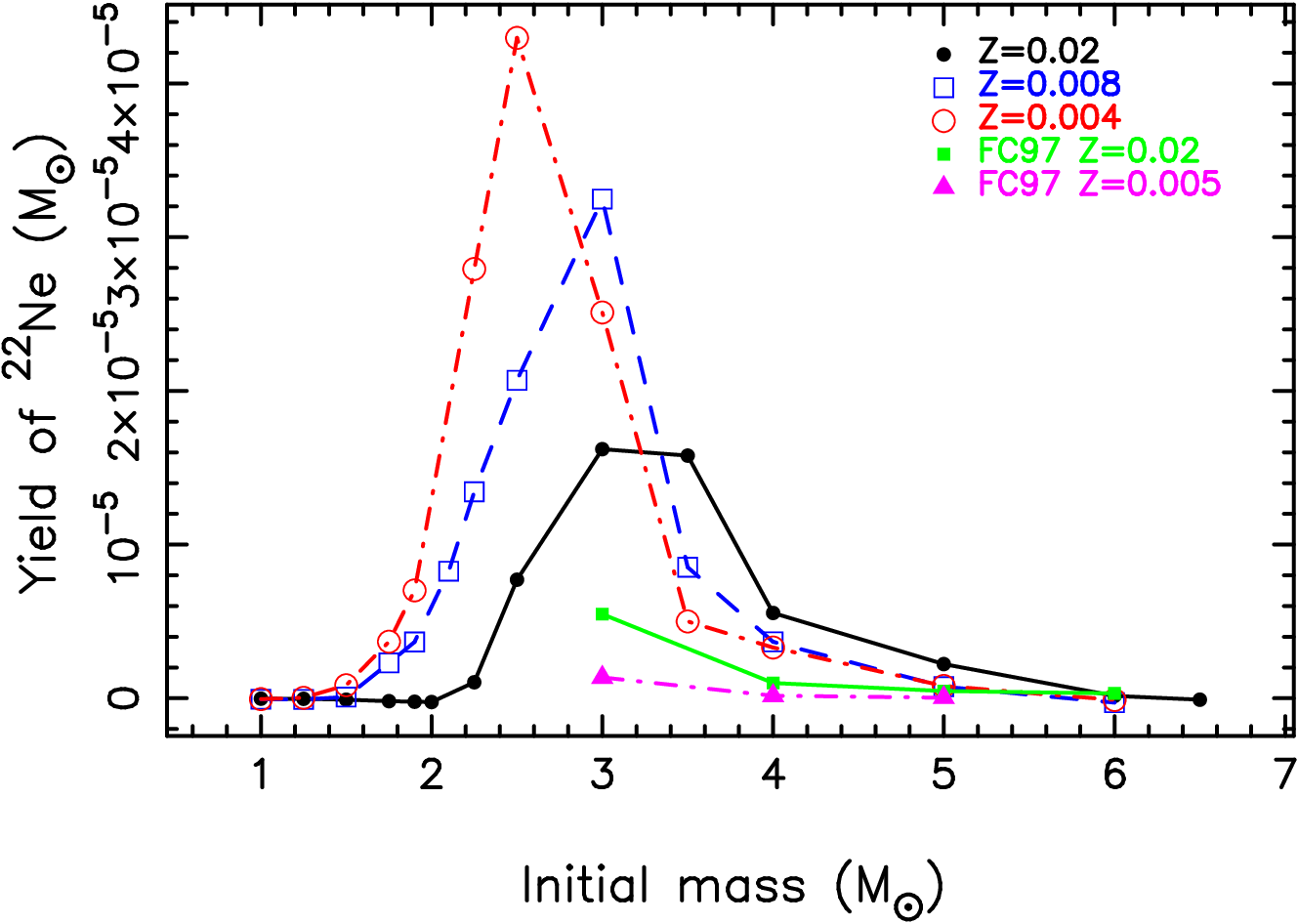} \\ 
\includegraphics[scale=0.5, angle=270]{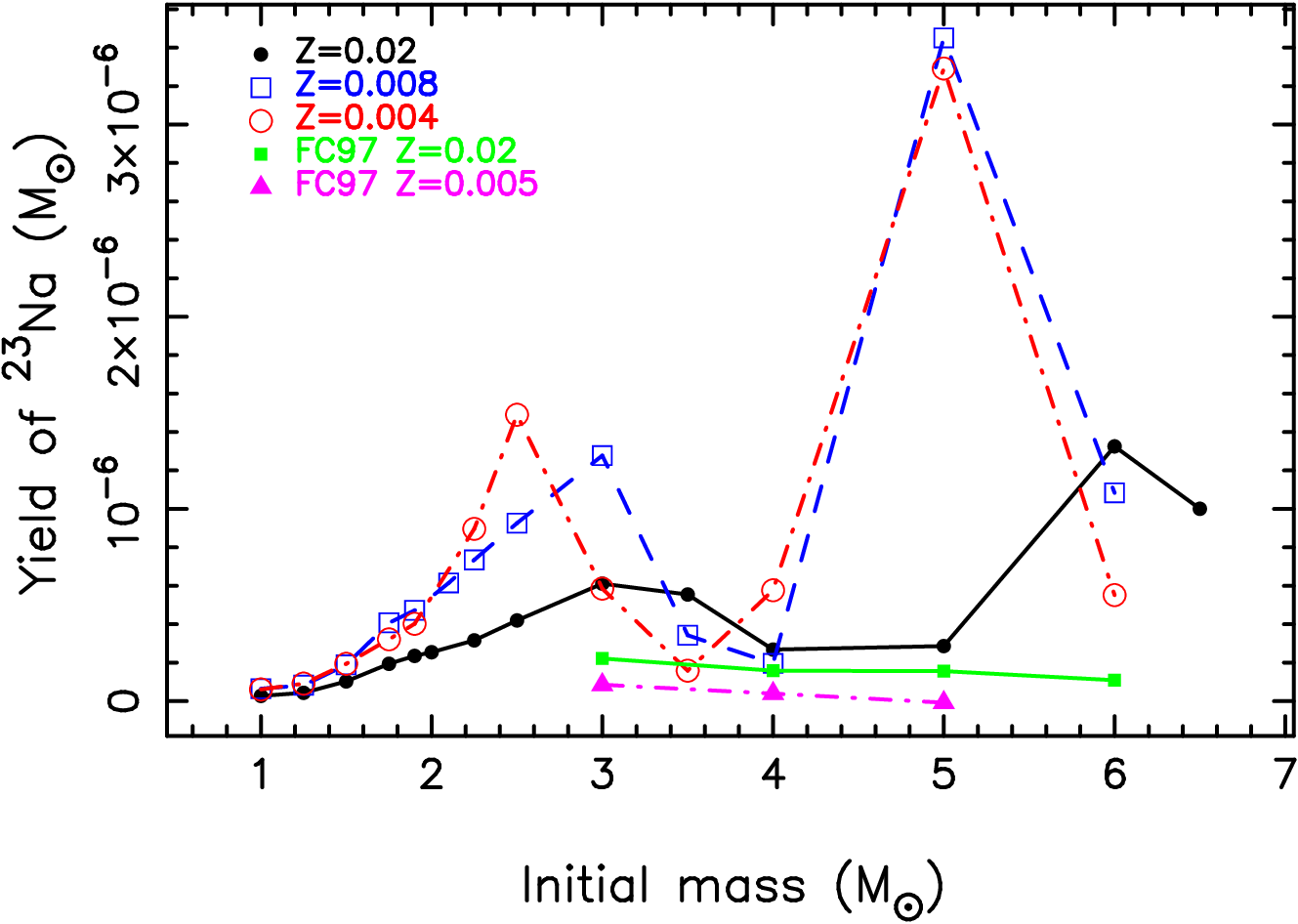} 
\end{tabular}
\caption{Weighted \iso{22}Ne (top panel) and \iso{23}Na 
(lower panel) yields  and for the $Z=0.02$, the $Z=0.008$ 
and $Z=0.004$ models. The yields of \citet{forestini97} 
are also shown.
}\label{ne22-na23}
\end{center}
\end{figure}

\begin{figure}
\begin{center}
\begin{tabular}{c}
\includegraphics[scale=0.5, angle=270]{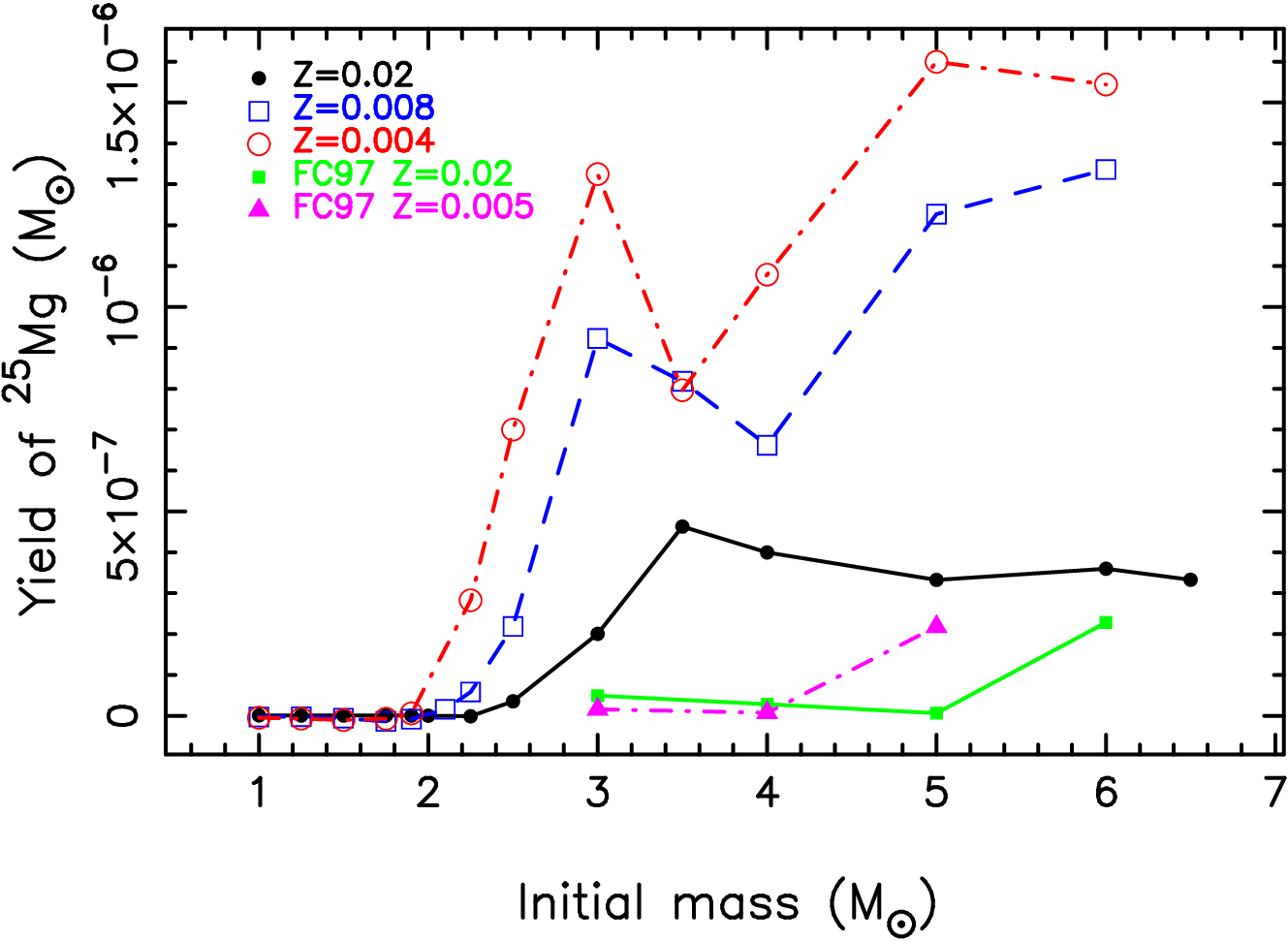} \\ 
\includegraphics[scale=0.5, angle=270]{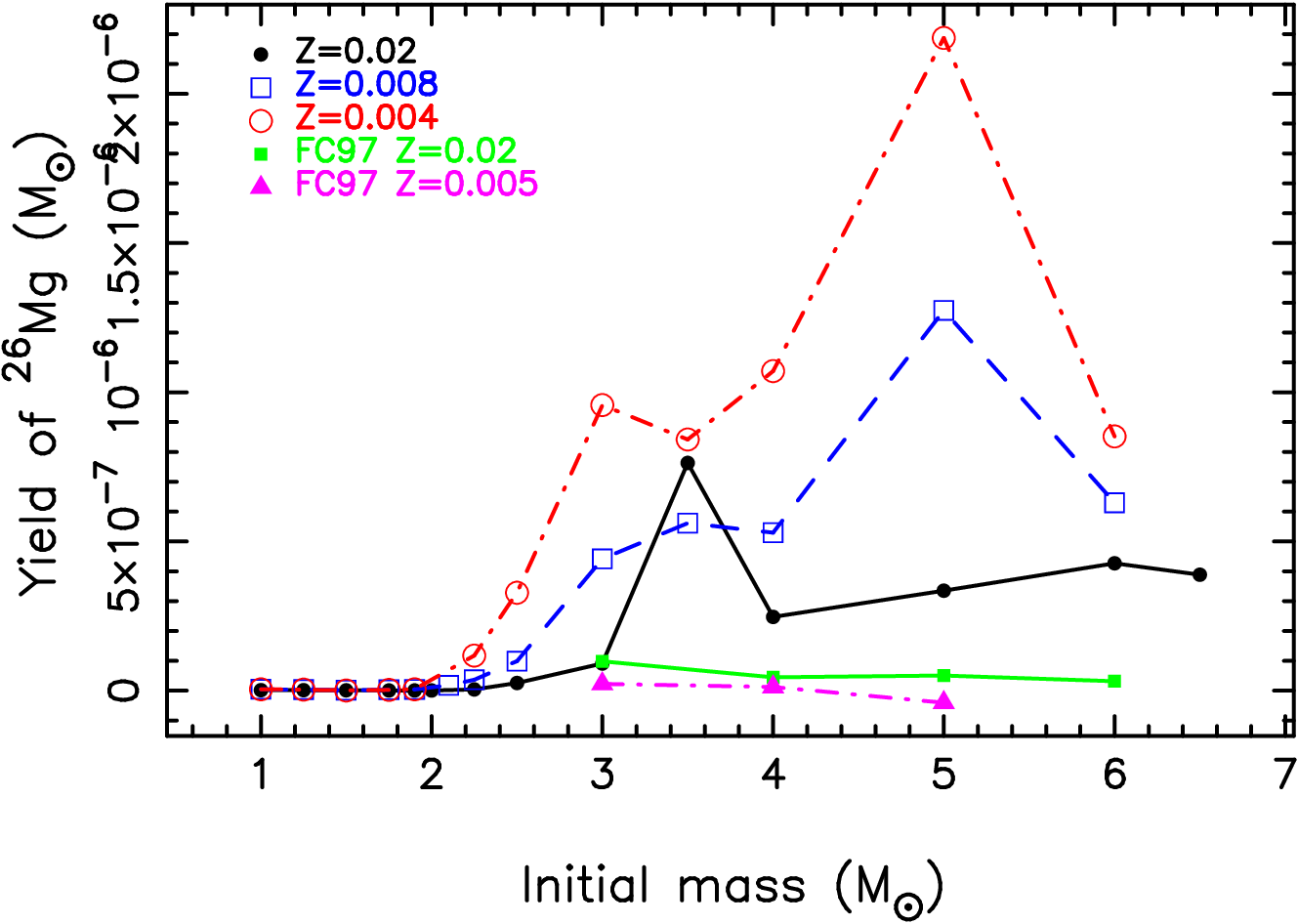} 
\end{tabular}
\caption{Weighted \iso{25}Mg (top panel) and \iso{26}Mg
(lower panel) yields  and for the $Z=0.02$, the $Z=0.008$ 
and $Z=0.004$ models. The yields of \citet{forestini97} 
are also shown.
}\label{mg25-mg26}
\end{center}
\end{figure}

\begin{figure}
\begin{center}
\begin{tabular}{c}
\includegraphics[scale=0.5, angle=270]{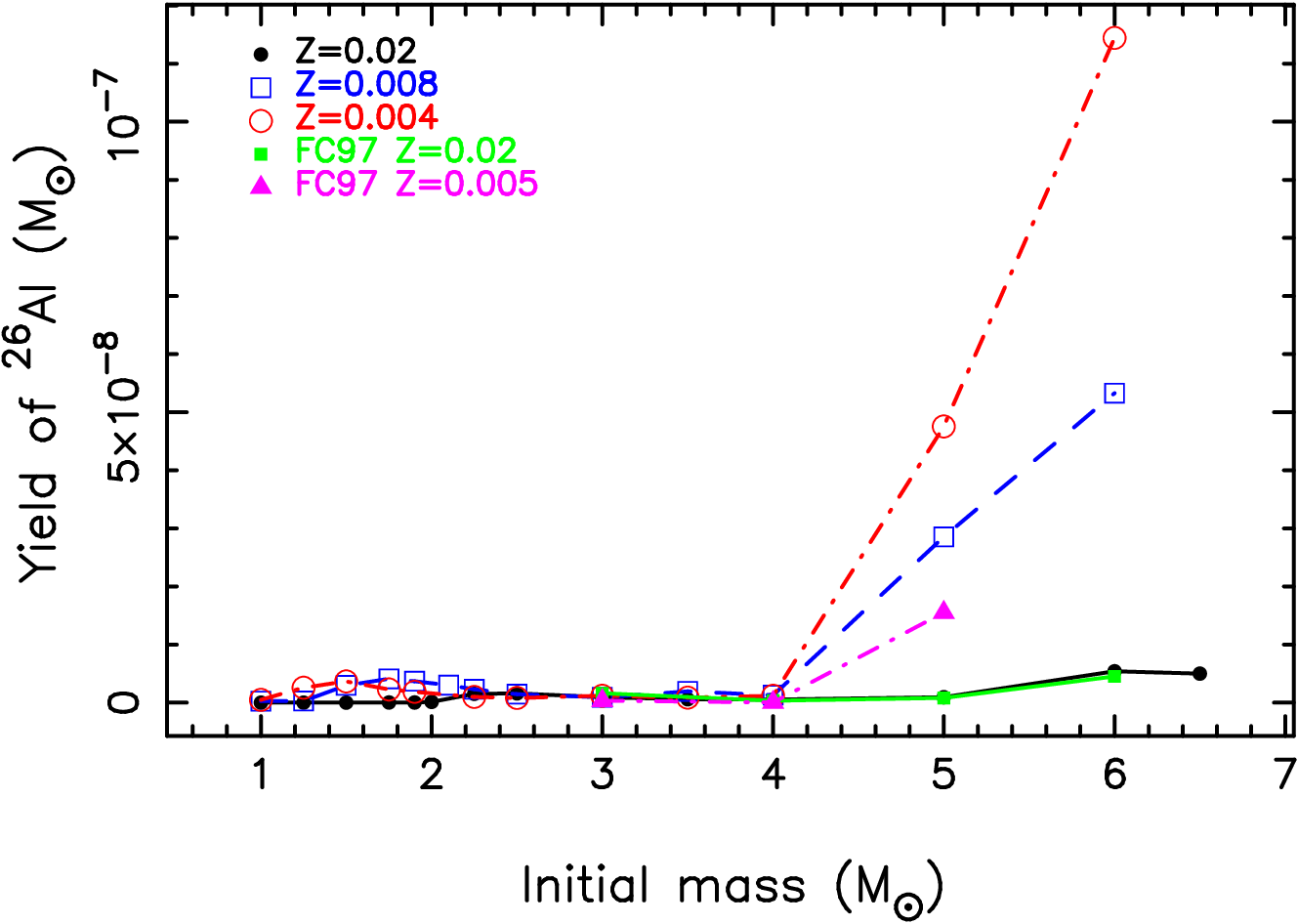} \\ 
\includegraphics[scale=0.5, angle=270]{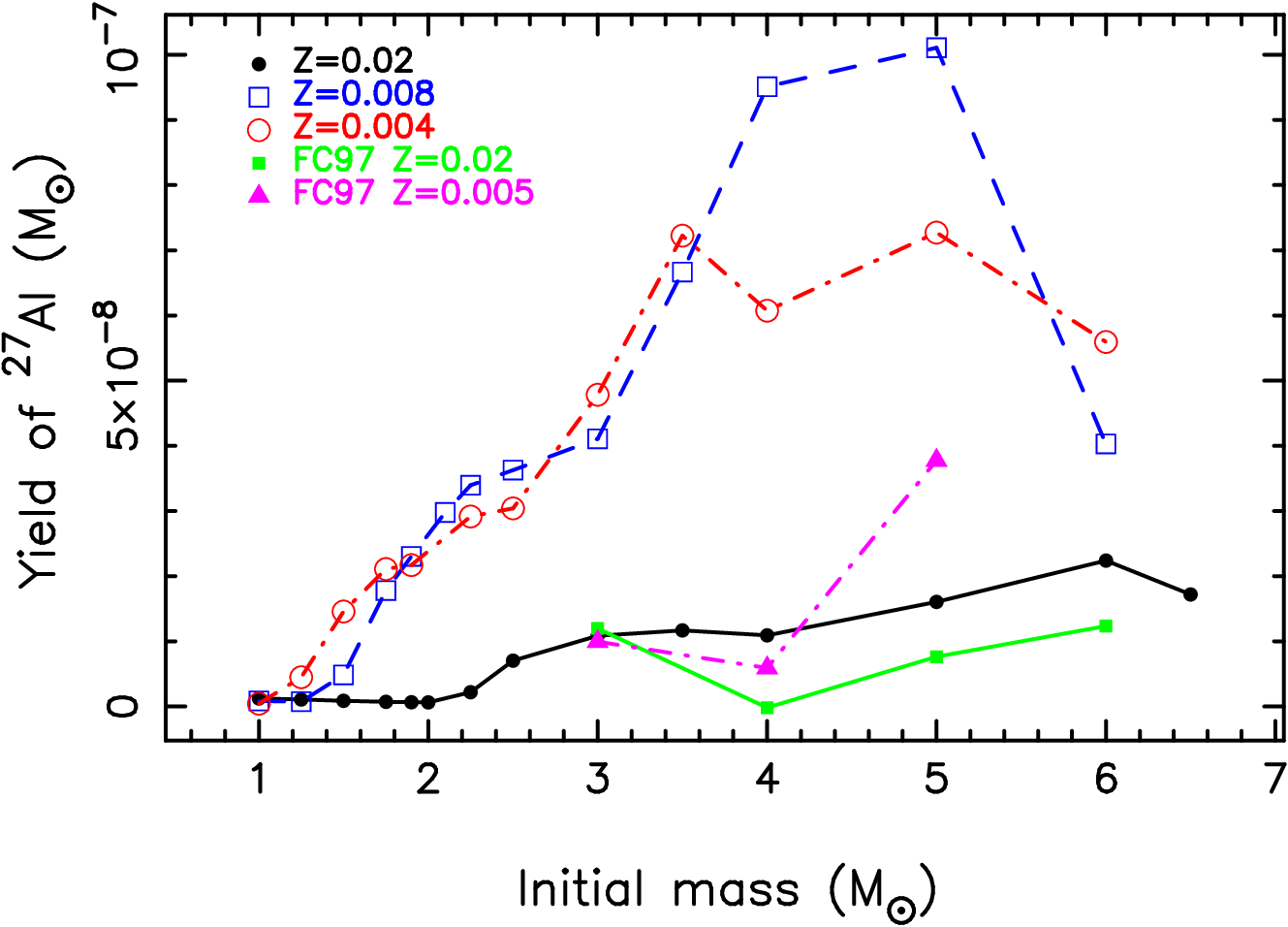} 
\end{tabular}
\caption{Weighted \iso{26}Al (top panel) and \iso{27}Al
(lower panel) yields  and for the $Z=0.02$, the $Z=0.008$ 
and $Z=0.004$ models. The yields of \citet{forestini97} 
are also shown.
}\label{al}
\end{center}
\end{figure}

In this section we provide a brief comparison of our yields 
to other authors. \citet{karakasThesis} and \citet{izzard04b} 
provided a detailed comparison of Izzard's synthetic yields to 
our detailed models (which his synthetic AGB algorithm is based 
upon), and to those from other synthetic AGB models
\citep{vandenhoek97,marigo01}. We do not repeat that discussion 
here but instead we present diagrams in Figs.~\ref{hyd} to~\ref{al},
and we discuss results for \iso{12}C and \iso{14}N. In Fig.~\ref{c12} 
we show the stellar yields of \iso{12}C, a nuclei representative of 
low-mass AGB evolution where the effects of TDU episodes are
important. In Fig.~\ref{n14} we show the yields of 
\iso{14}N, a nuclei that tracks HBB nucleosynthesis. In each
figure we show the results for three metallicities, $Z=0.02$,
0.008 and 0.004 compared to yields from synthetic AGB models, see 
the figure captions for more details. In each figure we have
weighted the yields by the three-component IMF of \citet{kroupa93}
such that
\begin{equation}
  y_{\rm i} = \frac{dY}{dM} = \xi(M_{0}) \, M_{\rm i},
 \label{eq:weightedyield}
\end{equation}
where $y_{\rm i}$ is the weighted yield and $M_{\rm i}$ is 
given by equation~\ref{eq:netyield}.  

From the Figs.~\ref{c12} and~\ref{n14} we see that our yields are 
almost the same as \citet{izzard04b} owing to the fact that these
synthetic models were tweaked to match our results. Our yields are
similar in behaviour to those of \citet{marigo01}, although because 
her models experience deeper TDU at a lower core mass, the yields of 
He-shell material e.g. \iso{4}He, \iso{12}C are higher. We notice
significant differences with \citet{vandenhoek97} especially in 
regards to \iso{14}N. This is owing to their simplistic treatment
of HBB nucleosynthesis which under-predicts the amount of CNO cycling
compared to the other computations. \citet{stancliffe07} provided a 
detailed comparison between yields from 1.5$\Msun$ models computed 
with the {\tt STARS} code to our yields, with the summary 
that their models experience deeper TDU and hence go in a 
similar direction to Marigo's, that is larger yields of material 
that is dredged from the He-burning shell.

In Fig.~\ref{f19} we show the (unweighted) yields of \iso{15}N
and \iso{19}F as a function of the initial mass and metallicity.
In this figure we include the $Z = 10^{-4}$ to show the effect 
of lower metallicity stellar evolution on the yields.  The 
maximum \iso{19}F yield is pushed to lower mass, owing to a 
higher He-burning shell and deeper TDU.  The yields of \iso{15}N
are negative, indicating this isotope is destroyed, and 
scale with metallicity except in the intermediate-mass 
$Z = 10^{-4}$ models, where we find that \iso{15}N is produced.
This is as a result of breakout from the \iso{14}N($p, \gamma$)\iso{15}O 
reaction at the temperatures and densities of HBB in these lower 
metallicity models. The production of \iso{15}N at lower metallicities
may have important consequences for the evolution of \iso{15}N
in the interstellar medium, especially since AGB stars are not
considered producers of this rare isotope.

We compare the $Z = 10^{-4}$ models to the AGB models
of \citet{herwig04b} of the same metallicity. This comparison is
made easier by the fact that both sets of models assumed a 
scaled-solar composition. Different mass loss was
assumed where we used VW93 mass loss and Herwig applied 
the \cite{bloecker95} prescription with the parameter 
$\eta_{\rm B} = 0.1$, that resulted in more rapid mass loss
than we observed. This can be demonstrated by considering
the 2$\Msun$, $Z= 10^{-4}$ model as an example. Our 2$\Msun$ 
calculation experienced 26 TPs compared to 7 for Herwig's model. 
Even though the maximum amount of matter dredged into the
envelope at any one TP was almost the same (1.112$\times 10^{-2}$ 
compared with Herwig's 1.171$\times 10^{-2}\Msun$), the total
amount of material dredged into the envelope over the
whole TP-AGB was much larger in our case (0.2183$\Msun$)
compared to Herwig's model (5.038$\times 10^{-2}\Msun$).
For this reason our yields and average surface mass fractions
are, in general, larger. The average abundance of \iso{12}C
in the wind of our model was $\sim 2$ times larger, of 
\iso{22}Ne 13 times larger and of \iso{23}Na 40 times larger.
However the average abundance of \iso{16}O was about 8
times larger in Herwig's model. These differences can be
attributed to 1) more TPs in our model, leading to more
TDU episodes and hotter nucleosynthesis conditions in
the He-shell, and 2) different intershell compositions 
caused by diffusive convective overshoot used in the 
Herwig model. Overshoot applied to the border at the
base of the convective pocket during a TP mixes some CO core
matter into the intershell, resulting in increased \iso{12}C 
and \iso{16}O abundances. These yields differences are similar
for all masses in the range 2 -- 6$\Msun$, where the
magnitude of the difference increases with increasing mass 
owing to the greater number of TPs experienced by our 
intermediate mass models (4, 5 and 6$\Msun$).

\section{Conclusions}

We present stellar structure data and yields from detailed 
AGB models with initial masses between 1 to 6$\Msun$, and with 
metallicities in the range from solar to $1/200^{\rm th}$ of the 
solar metallicity. All models were evolved from the zero-age main
sequence to near the tip of the AGB track, from there we used
a synthetic AGB algorithm to estimate the contribution of
any remaining TPs.   All results are 
presented in tabulated form and are available for download. 
These data are useful for many applications including 
comparison to other detailed AGB models, for use in galactic chemical 
evolution studies and for comparison to the composition of 
AGB stars, planetary nebulae and other post-AGB objects, 
as well as pre-solar grains.  Such a comparison may help shed 
light on the model uncertainties that affect the yields, 
such as the mass-loss rate or nuclear reaction rates.
Future work will involve running scaled-solar models for
the intermediate-metallicities ($Z = 0.008, 0.004$) and  
adding more nuclei in the post-processing nuclear network 
to enable us to provide yields for $s$-process elements. 
Future work should also look at some of the reasons for
convergence problems in the stellar models near the tip of
the AGB, and include more physics into the calculations 
such as extra-mixing in low-mass AGB stars.

\section*{Acknowledgments} 

AIK gratefully acknowledges support from the Australian Research 
Council's Discovery Projects funding scheme (project number DP0664105).
JCL thanks the ARC for support. We also thank the referee, Carlos Abia,
for instructive comments that have helped to improve the manuscript.

\bibliography{mnemonic,/Users/amanda/biblio/library}

\begin{thebibliography}{86}
\expandafter\ifx\csname natexlab\endcsname\relax\def\natexlab#1{#1}\fi

\bibitem[{{Abia} \& {Isern}(1997)}]{abia97}
{Abia}, C. \& {Isern}, J. 1997, MNRAS, 289, L11

\bibitem[{{Anders} \& {Grevesse}(1989)}]{anders89}
{Anders}, E. \& {Grevesse}, N. 1989, Geochim. Cosmochim. Acta, 53, 197

\bibitem[{{Asplund} {et~al.}(2005){Asplund}, {Grevesse}, \&
  {Sauval}}]{asplund05}
{Asplund}, M., {Grevesse}, N., \& {Sauval}, A.~J. 2005, in ASP Conf. Ser. 336:
  Cosmic Abundances as Records of Stellar Evolution and Nucleosynthesis, ed.
  T.~G. {Barnes}, III \& F.~N. {Bash}, 25

\bibitem[{{Beers} \& {Christlieb}(2005)}]{beers05}
{Beers}, T.~C. \& {Christlieb}, N. 2005, ARA\&A, 43, 531

\bibitem[{{Bl{\"o}cker}(1995)}]{bloecker95}
{Bl{\"o}cker}, T. 1995, A\&A, 297, 727

\bibitem[{{Bloecker} \& {Schoenberner}(1991)}]{bloecker91}
{Bloecker}, T. \& {Schoenberner}, D. 1991, A\&A, 244, L43

\bibitem[{{Boothroyd} \& {Sackmann}(1988)}]{boothroyd88c}
{Boothroyd}, A.~I. \& {Sackmann}, I.-J. 1988, ApJ, 328, 653

\bibitem[{{Boothroyd} {et~al.}(1993){Boothroyd}, {Sackmann}, \&
  {Ahern}}]{boothroyd93}
{Boothroyd}, A.~I., {Sackmann}, I.-J., \& {Ahern}, S.~C. 1993, ApJ, 416, 762

\bibitem[{{Boothroyd} {et~al.}(1995){Boothroyd}, {Sackmann}, \&
  {Wasserburg}}]{boothroyd95}
{Boothroyd}, A.~I., {Sackmann}, I.-J., \& {Wasserburg}, G.~J. 1995, ApJ, 442,
  L21

\bibitem[{{Busso} {et~al.}(1999){Busso}, {Gallino}, \& {Wasserburg}}]{busso99}
{Busso}, M., {Gallino}, R., \& {Wasserburg}, G.~J. 1999, ARA\&A, 37, 239

\bibitem[{{Campbell}(2007)}]{campbellThesis}
{Campbell}, S.~W. 2007, PhD thesis, Monash University

\bibitem[{{Cannon}(1993)}]{cannon93}
{Cannon}, R.~C. 1993, MNRAS, 263, 817

\bibitem[{{Charbonnel}(1994)}]{charbonnel94}
{Charbonnel}, C. 1994, A\&A, 282, 811

\bibitem[{{Charbonnel} \& {Zahn}(2007)}]{charbonnel07}
{Charbonnel}, C. \& {Zahn}, J.-P. 2007, A\&A, 467, L15

\bibitem[{{Dearborn} {et~al.}(2006){Dearborn}, {Lattanzio}, \&
  {Eggleton}}]{dearborn06}
{Dearborn}, D.~S.~P., {Lattanzio}, J.~C., \& {Eggleton}, P.~P. 2006, ApJ, 639,
  405

\bibitem[{{Denissenkov} \& {Herwig}(2003)}]{denissenkov03b}
{Denissenkov}, P.~A. \& {Herwig}, F. 2003, ApJ, 590, L99

\bibitem[{{Dinerstein} {et~al.}(2003){Dinerstein}, {Richter}, {Lacy}, \&
  {Sellgren}}]{dinerstein03}
{Dinerstein}, H.~L., {Richter}, M.~J., {Lacy}, J.~H., \& {Sellgren}, K. 2003,
  AJ, 125, 265

\bibitem[{{Dopita} {et~al.}(1997){Dopita}, {Vassiliadis}, {Wood},
  {Meatheringham}, {Harrington}, {Bohlin}, {Ford}, {Stecher}, \&
  {Maran}}]{dopita97}
{Dopita}, M.~A., {Vassiliadis}, E., {Wood}, P.~R., {Meatheringham}, S.~J.,
  {Harrington}, J.~P., {Bohlin}, R.~C., {Ford}, H.~C., {Stecher}, T.~P., \&
  {Maran}, S.~P. 1997, ApJ, 474, 188

\bibitem[{{Eggleton} {et~al.}(2006){Eggleton}, {Dearborn}, \&
  {Lattanzio}}]{eggleton06}
{Eggleton}, P.~P., {Dearborn}, D.~S.~P., \& {Lattanzio}, J.~C. 2006, Science,
  314, 1580

\bibitem[{{Eggleton} {et~al.}(2007){Eggleton}, {Dearborn}, \&
  {Lattanzio}}]{eggleton07}
---. 2007, ApJ, submitted

\bibitem[{{Fenner} {et~al.}(2004){Fenner}, {Campbell}, {Karakas}, {Lattanzio},
  \& {Gibson}}]{fenner04}
{Fenner}, Y., {Campbell}, S., {Karakas}, A.~I., {Lattanzio}, J.~C., \&
  {Gibson}, B.~K. 2004, MNRAS, 353, 789

\bibitem[{{Forestini} \& {Charbonnel}(1997)}]{forestini97}
{Forestini}, M. \& {Charbonnel}, C. 1997, A\&AS, 123, 241

\bibitem[{{Frogel} {et~al.}(1990){Frogel}, {Mould}, \& {Blanco}}]{frogel90}
{Frogel}, J.~A., {Mould}, J., \& {Blanco}, V.~M. 1990, ApJ, 352, 96

\bibitem[{{Frost} {et~al.}(1998){Frost}, {Cannon}, {Lattanzio}, {Wood}, \&
  {Forestini}}]{frost98a}
{Frost}, C.~A., {Cannon}, R.~C., {Lattanzio}, J.~C., {Wood}, P.~R., \&
  {Forestini}, M. 1998, A\&A, 332, L17

\bibitem[{{Frost} \& {Lattanzio}(1996)}]{frost96}
{Frost}, C.~A. \& {Lattanzio}, J.~C. 1996, ApJ, 473, 383

\bibitem[{{Garcia-Hernandez} {et~al.}(2006){Garcia-Hernandez}, {Garcia-Lario},
  {Plez}, {D'Antona}, {Manchado}, \& {Trigo-Rodriguez}}]{garcia06}
{Garcia-Hernandez}, D.~A., {Garcia-Lario}, P., {Plez}, B., {D'Antona}, F.,
  {Manchado}, A., \& {Trigo-Rodriguez}, J.~M. 2006, Science, 314, 1751

\bibitem[{{Garcia-Hernandez} {et~al.}(2007){Garcia-Hernandez}, {Garcia-Lario},
  {Plez}, {Manchado}, {D'Antona}, {Lub}, \& {Habing}}]{garcia07}
{Garcia-Hernandez}, D.~A., {Garcia-Lario}, P., {Plez}, B., {Manchado}, A.,
  {D'Antona}, F., {Lub}, J., \& {Habing}, H. 2007, A\&A, 462, 711

\bibitem[{{Goriely} \& {Mowlavi}(2000)}]{goriely00}
{Goriely}, S. \& {Mowlavi}, N. 2000, A\&A, 362, 599

\bibitem[{{Gratton} {et~al.}(2004){Gratton}, {Sneden}, \&
  {Carretta}}]{gratton04}
{Gratton}, R., {Sneden}, C., \& {Carretta}, E. 2004, ARA\&A, 42, 385

\bibitem[{{Groenewegen} \& {de Jong}(1993)}]{groen93}
{Groenewegen}, M.~A.~T. \& {de Jong}, T. 1993, A\&A, 267, 410

\bibitem[{{Guandalini} {et~al.}(2006){Guandalini}, {Busso}, {Ciprini},
  {Silvestro}, \& {Persi}}]{guandalini06}
{Guandalini}, R., {Busso}, M., {Ciprini}, S., {Silvestro}, G., \& {Persi}, P.
  2006, A\&A, 445, 1069

\bibitem[{{Henry}(1989)}]{henry89}
{Henry}, R.~B.~C. 1989, MNRAS, 241, 453

\bibitem[{{Herwig}(2004{\natexlab{a}})}]{herwig04a}
{Herwig}, F. 2004{\natexlab{a}}, ApJ, 605, 425

\bibitem[{{Herwig}(2004{\natexlab{b}})}]{herwig04b}
---. 2004{\natexlab{b}}, ApJS, 155, 651

\bibitem[{{Herwig}(2005)}]{herwig05}
---. 2005, ARA\&A, 43, 435

\bibitem[{{Izzard} {et~al.}(2006){Izzard}, {Karakas}, {Lugaro}, \&
  {Tout}}]{izzard06}
{Izzard}, R.~G.and~{Dray}, L.~M., {Karakas}, A.~I., {Lugaro}, M., \& {Tout},
  C.~A. 2006, A\&A, 460, 565

\bibitem[{{Izzard} {et~al.}(2007){Izzard}, {Lugaro}, {Karakas}, {Iliadis}, \&
  {van Raai}}]{izzard07}
{Izzard}, R.~G., {Lugaro}, M., {Karakas}, A.~I., {Iliadis}, C., \& {van Raai},
  M. 2007, A\&A, 466, 641

\bibitem[{{Izzard} {et~al.}(2004){Izzard}, {Tout}, {Karakas}, \&
  {Pols}}]{izzard04b}
{Izzard}, R.~G., {Tout}, C.~A., {Karakas}, A.~I., \& {Pols}, O.~R. 2004, MNRAS,
  350, 407

\bibitem[{{Jorissen} {et~al.}(1992){Jorissen}, {Smith}, \&
  {Lambert}}]{jorissen92}
{Jorissen}, A., {Smith}, V.~V., \& {Lambert}, D.~L. 1992, A\&A, 261, 164

\bibitem[{{Kaler}(1978)}]{kaler78}
{Kaler}, J.~B. 1978, ApJ, 225, 527

\bibitem[{{Karakas}(2003)}]{karakasThesis}
{Karakas}, A.~I. 2003, PhD thesis, Monash University

\bibitem[{{Karakas} {et~al.}(2006{\natexlab{a}}){Karakas}, {Fenner}, {Sills},
  {Campbell}, \& {Lattanzio}}]{karakas06b}
{Karakas}, A.~I., {Fenner}, Y., {Sills}, A., {Campbell}, S.~W., \& {Lattanzio},
  J.~C. 2006{\natexlab{a}}, ApJ, 652, 1240

\bibitem[{{Karakas} \& {Lattanzio}(2003)}]{karakas03b}
{Karakas}, A.~I. \& {Lattanzio}, J.~C. 2003, PASA, 20, 279

\bibitem[{{Karakas} {et~al.}(2002){Karakas}, {Lattanzio}, \&
  {Pols}}]{karakas02}
{Karakas}, A.~I., {Lattanzio}, J.~C., \& {Pols}, O.~R. 2002, PASA, 19, 515

\bibitem[{{Karakas} {et~al.}(2007){Karakas}, {Lugaro}, \&
  {Gallino}}]{karakas07a}
{Karakas}, A.~I., {Lugaro}, M., \& {Gallino}, R. 2007, ApJ, 656, L73

\bibitem[{{Karakas} {et~al.}(2006{\natexlab{b}}){Karakas}, {Lugaro},
  {Wiescher}, {Goerres}, \& {Ugalde}}]{karakas06a}
{Karakas}, A.~I., {Lugaro}, M., {Wiescher}, M., {Goerres}, J., \& {Ugalde}, C.
  2006{\natexlab{b}}, ApJ, 643, 471

\bibitem[{{Kroupa} {et~al.}(1993){Kroupa}, {Tout}, \& {Gilmore}}]{kroupa93}
{Kroupa}, P., {Tout}, C.~A., \& {Gilmore}, G. 1993, MNRAS, 262, 545

\bibitem[{{Lattanzio} {et~al.}(1996){Lattanzio}, {Frost}, {Cannon}, \&
  {Wood}}]{lattanzio96}
{Lattanzio}, J., {Frost}, C., {Cannon}, R., \& {Wood}, P.~R. 1996, Mem. Soc.
  Astron. Italiana, 67, 729

\bibitem[{{Lattanzio}(1986)}]{lattanzio86}
{Lattanzio}, J.~C. 1986, ApJ, 311, 708

\bibitem[{{Lattanzio}(1992)}]{lattanzio92}
---. 1992, PASA, 10, 120

\bibitem[{{Lugaro} {et~al.}(2007){Lugaro}, {Karakas}, {Nittler}, {Alexander},
  {Hoppe}, {Iliadis}, \& {Lattanzio}}]{lugaro07}
{Lugaro}, M., {Karakas}, A.~I., {Nittler}, L.~R., {Alexander}, C.~M.~O.,
  {Hoppe}, P., {Iliadis}, C., \& {Lattanzio}, J.~C. 2007, A\&A, 461, 657

\bibitem[{{Lugaro} {et~al.}(2004){Lugaro}, {Ugalde}, {Karakas}, {G{\"o}rres},
  {Wiescher}, {Lattanzio}, \& {Cannon}}]{lugaro04}
{Lugaro}, M., {Ugalde}, C., {Karakas}, A.~I., {G{\"o}rres}, J., {Wiescher}, M.,
  {Lattanzio}, J.~C., \& {Cannon}, R.~C. 2004, ApJ, 615, 934

\bibitem[{{Marigo}(2001)}]{marigo01}
{Marigo}, P. 2001, A\&A, 370, 194

\bibitem[{{Marigo} {et~al.}(1996){Marigo}, {Bressan}, \& {Chiosi}}]{marigo96}
{Marigo}, P., {Bressan}, A., \& {Chiosi}, C. 1996, A\&A, 313, 545

\bibitem[{{McSaveney} {et~al.}(2007){McSaveney}, {Wood}, {Scholz}, {Lattanzio},
  \& {Hinkle}}]{mcsaveney07}
{McSaveney}, J.~A., {Wood}, P.~R., {Scholz}, M., {Lattanzio}, J.~C., \&
  {Hinkle}, K.~H. 2007, MNRAS, in press

\bibitem[{{Mowlavi}(1999)}]{mowlavi99a}
{Mowlavi}, N. 1999, A\&A, 344, 617

\bibitem[{{Nollett} {et~al.}(2003){Nollett}, {Busso}, \&
  {Wasserburg}}]{nollett03}
{Nollett}, K.~M., {Busso}, M., \& {Wasserburg}, G.~J. 2003, ApJ, 582, 1036

\bibitem[{{Paczynski}(1975)}]{paczynski75}
{Paczynski}, B. 1975, ApJ, 202, 558

\bibitem[{{Palacios} {et~al.}(2006){Palacios}, {Charbonnel}, {Talon}, \&
  {Siess}}]{palacios06}
{Palacios}, A., {Charbonnel}, C., {Talon}, S., \& {Siess}, L. 2006, A\&A, 453,
  261

\bibitem[{{Plez} {et~al.}(1993){Plez}, {Smith}, \& {Lambert}}]{plez93}
{Plez}, B., {Smith}, V.~V., \& {Lambert}, D.~L. 1993, ApJ, 418, 812

\bibitem[{{Reimers}(1975)}]{reimers75}
{Reimers}, D. 1975, {Circumstellar envelopes and mass loss of red giant stars}
  (Problems in stellar atmospheres and envelopes.), 229--256

\bibitem[{{Renzini} \& {Voli}(1981)}]{renzini81}
{Renzini}, A. \& {Voli}, M. 1981, A\&A, 94, 175

\bibitem[{{Russell} \& {Dopita}(1992)}]{russell92}
{Russell}, S.~C. \& {Dopita}, M.~A. 1992, ApJ, 384, 508

\bibitem[{{Sharpee} {et~al.}(2007){Sharpee}, {Zhang}, {Williams}, {Pellegrini},
  {Cavagnolo}, {Baldwin}, {Phillips}, \& {Liu}}]{sharpee07}
{Sharpee}, B., {Zhang}, Y., {Williams}, R., {Pellegrini}, E., {Cavagnolo}, K.,
  {Baldwin}, J.~A., {Phillips}, M., \& {Liu}, X.-W. 2007, ApJ, 659, 1265

\bibitem[{{Siess}(2006)}]{siess06}
{Siess}, L. 2006, A\&A, 448, 717

\bibitem[{{Smith} \& {Lambert}(1989)}]{smith89}
{Smith}, V.~V. \& {Lambert}, D.~L. 1989, ApJ, 345, L75

\bibitem[{{Smith} \& {Lambert}(1990)}]{smith90b}
---. 1990, ApJ, 361, L69

\bibitem[{{Stancliffe} {et~al.}(2005){Stancliffe}, {Izzard}, \&
  {Tout}}]{stancliffe05a}
{Stancliffe}, R.~J., {Izzard}, R.~G., \& {Tout}, C.~A. 2005, MNRAS, 356, L1

\bibitem[{{Stancliffe} {et~al.}(2004){Stancliffe}, {Izzard}, {Tout}, \&
  {Pols}}]{stancliffe04a}
{Stancliffe}, R.~J., {Izzard}, R.~G., {Tout}, C.~A., \& {Pols}, O.~R. 2004,
  Mem. Soc. Astron. Italiana, 75, 670

\bibitem[{{Stancliffe} \& {Jeffery}(2007)}]{stancliffe07}
{Stancliffe}, R.~J. \& {Jeffery}, C.~S. 2007, MNRAS, 375, 1280

\bibitem[{{Stanghellini} {et~al.}(2006){Stanghellini}, {Guerrero}, {Cunha},
  {Manchado}, \& {Villaver}}]{stanghellini06}
{Stanghellini}, L., {Guerrero}, M.~A., {Cunha}, K., {Manchado}, A., \&
  {Villaver}, E. 2006, ApJ, 651, 898

\bibitem[{{Stanghellini} {et~al.}(2000){Stanghellini}, {Shaw}, {Balick}, \&
  {Blades}}]{stanghellini00}
{Stanghellini}, L., {Shaw}, R.~A., {Balick}, B., \& {Blades}, J.~C. 2000, ApJ,
  534, L167

\bibitem[{{Sterling} \& {Dinerstein}(2007)}]{sterling07b}
{Sterling}, N.~C. \& {Dinerstein}. 2007, ApJS, accepted

\bibitem[{{Sterling} {et~al.}(2002){Sterling}, {Dinerstein}, \&
  {Bowers}}]{sterling02}
{Sterling}, N.~C., {Dinerstein}, H.~L., \& {Bowers}, C.~W. 2002, ApJ, 578, L55

\bibitem[{{Straniero} {et~al.}(1997){Straniero}, {Chieffi}, {Limongi}, {Busso},
  {Gallino}, \& {Arlandini}}]{straniero97}
{Straniero}, O., {Chieffi}, A., {Limongi}, M., {Busso}, M., {Gallino}, R., \&
  {Arlandini}, C. 1997, ApJ, 478, 332

\bibitem[{{Thielemann} {et~al.}(1986){Thielemann}, {Truran}, \&
  {Arnould}}]{thielemann86}
{Thielemann}, F.-K., {Truran}, J.~W., \& {Arnould}, M. 1986, in Advances in
  Nuclear Astrophysics, ed. E.~{Vangioni-Flam}, J.~{Audouze}, M.~{Casse}, J.-P.
  {Chieze}, \& J.~{Tran Thanh van}, 525--540

\bibitem[{{van den Hoek} \& {Groenewegen}(1997)}]{vandenhoek97}
{van den Hoek}, L.~B. \& {Groenewegen}, M.~A.~T. 1997, A\&AS, 123, 305

\bibitem[{{van Winckel}(2003)}]{vanwinckel03}
{van Winckel}, H. 2003, ARA\&A, 41, 391

\bibitem[{{Vassiliadis} \& {Wood}(1993)}]{vw93}
{Vassiliadis}, E. \& {Wood}, P.~R. 1993, ApJ, 413, 641

\bibitem[{{Ventura} \& {D'Antona}(2005{\natexlab{a}})}]{ventura05a}
{Ventura}, P. \& {D'Antona}, F. 2005{\natexlab{a}}, A\&A, 431, 279

\bibitem[{{Ventura} \& {D'Antona}(2005{\natexlab{b}})}]{ventura05b}
---. 2005{\natexlab{b}}, A\&A, 439, 1075

\bibitem[{{Ventura} {et~al.}(2002){Ventura}, {D'Antona}, \&
  {Mazzitelli}}]{ventura02}
{Ventura}, P., {D'Antona}, F., \& {Mazzitelli}, I. 2002, A\&A, 393, 215

\bibitem[{{Wagenhuber} \& {Groenewegen}(1998)}]{wagenhuber98}
{Wagenhuber}, J. \& {Groenewegen}, M.~A.~T. 1998, A\&A, 340, 183

\bibitem[{{Wallerstein} \& {Knapp}(1998)}]{wallerstein98}
{Wallerstein}, G. \& {Knapp}, G.~R. 1998, ARA\&A, 36, 369

\bibitem[{{Wood} {et~al.}(1983){Wood}, {Bessell}, \& {Fox}}]{wood83}
{Wood}, P.~R., {Bessell}, M.~S., \& {Fox}, M.~W. 1983, ApJ, 272, 99

\bibitem[{{Yong} {et~al.}(2003){Yong}, {Grundahl}, {Lambert}, {Nissen}, \&
  {Shetrone}}]{yong03a}
{Yong}, D., {Grundahl}, F., {Lambert}, D.~L., {Nissen}, P.~E., \& {Shetrone},
  M.~D. 2003, A\&A, 402, 985

\end{thebibliography}



\pagebreak

\begin{table}[t]
\begin{center}
\caption{Data for the 3$\Msun$, $Z = 0.02$ model.}\label{example1}
\begin{tabular}{ccccccc}
\hline Pulse & $M_{\rm core}$ & $M_{\rm csh}$ & $t_{\rm csh}$ &
 $\Delta M_{\rm dredge}$ & $\lambda$ & $\lambda_{\rm dup}$ \\
\hline 
\hline
  1 & 5.763726E-01 & 2.237684E-02 & 3.339341E+02 & 0.000000E+00 & 0.000000E+00 & 0.000000E+00 \\
  2 & 5.782992E-01 & 2.551377E-02 & 2.015471E+02 & 0.000000E+00 & 0.000000E+00 & 0.000000E+00 \\
  3 & 5.811490E-01 & 2.586406E-02 & 2.061110E+02 & 0.000000E+00 & 0.000000E+00 & 0.000000E+00 \\
  4 & 5.850938E-01 & 2.622330E-02 & 1.793678E+02 & 0.000000E+00 & 0.000000E+00 & 0.000000E+00 \\
  5 & 5.898175E-01 & 2.566504E-02 & 1.954225E+02 & 0.000000E+00 & 0.000000E+00 & 0.000000E+00 \\
  6 & 5.952812E-01 & 2.515078E-02 & 1.755154E+02 & 0.000000E+00 & 0.000000E+00 & 0.000000E+00 \\
  7 & 6.012007E-01 & 2.422720E-02 & 2.020128E+02 & 0.000000E+00 & 0.000000E+00 & 0.000000E+00 \\
  8 & 6.074721E-01 & 2.330387E-02 & 1.798662E+02 & 0.000000E+00 & 0.000000E+00 & 0.000000E+00 \\
  9 & 6.139326E-01 & 2.226830E-02 & 1.702774E+02 & 0.000000E+00 & 0.000000E+00 & 0.000000E+00 \\
 10 & 6.205002E-01 & 2.124983E-02 & 1.376081E+02 & 2.955794E-04 & 4.524140E-02 & 1.390973E-02 \\
 11 & 6.269240E-01 & 2.039099E-02 & 1.702900E+02 & 6.338358E-04 & 9.432908E-02 & 3.108412E-02 \\
 12 & 6.331832E-01 & 1.966518E-02 & 1.570454E+02 & 1.435995E-03 & 2.083254E-01 & 7.302221E-02 \\
 13 & 6.389904E-01 & 1.907504E-02 & 1.575242E+02 & 2.473533E-03 & 3.414993E-01 & 1.296738E-01 \\
 14 & 6.442509E-01 & 1.871735E-02 & 1.404630E+02 & 3.429413E-03 & 4.434169E-01 & 1.832211E-01 \\
 15 & 6.489711E-01 & 1.843846E-02 & 1.480161E+02 & 4.516482E-03 & 5.541952E-01 & 2.449490E-01 \\
 16 & 6.531649E-01 & 1.828820E-02 & 1.291233E+02 & 4.803419E-03 & 5.514702E-01 & 2.626514E-01 \\
 17 & 6.571379E-01 & 1.772785E-02 & 1.474301E+02 & 5.583823E-03 & 6.362296E-01 & 3.149746E-01 \\
 18 & 6.607848E-01 & 1.762992E-02 & 1.466641E+02 & 6.045818E-03 & 6.549662E-01 & 3.429294E-01 \\
 19 & 6.641064E-01 & 1.734108E-02 & 1.509221E+02 & 6.349206E-03 & 6.777933E-01 & 3.661368E-01 \\
 20 & 6.672819E-01 & 1.703131E-02 & 1.502906E+02 & 6.745815E-03 & 7.082441E-01 & 3.960831E-01 \\
 21 & 6.702475E-01 & 1.687807E-02 & 1.387330E+02 & 7.370591E-03 & 7.589639E-01 & 4.366964E-01 \\
 22 & 6.729269E-01 & 1.684576E-02 & 1.398080E+02 & 7.480979E-03 & 7.443761E-01 & 4.440867E-01 \\
 23 & 6.753748E-01 & 1.643026E-02 & 1.353630E+02 & 7.710814E-03 & 7.766045E-01 & 4.693057E-01 \\
 24 & 6.777956E-01 & 1.632255E-02 & 1.362144E+02 & 7.740200E-03 & 7.639663E-01 & 4.742027E-01 \\
 25 & 6.800572E-01 & 1.598930E-02 & 1.370616E+02 & 6.662667E-03 & 6.661482E-01 & 4.166952E-01 \\
\hline
\hline
\end{tabular}
\medskip\\
\end{center}
\end{table}

\begin{landscape}
\begin{table}[t]
\begin{center}
\caption{Data for the 3$\Msun$, $Z = 0.02$ model continued.}\label{example2}
\begin{tabular}{cccccccc}
\hline Pulse & $T_{\rm Heshell}$ & $T_{\rm bce}$ & $T_{\rm Hshell}$ & interpulse 
& $M_{\rm tot}$ &  MaxL & MaxLHe \\
\hline
\hline
 1 & 1.840110E+08 & 2.505874E+06 & 4.832516E+07 & 0.000000E+00 & 2.989960E+00 & 3.297096E+03 & 2.896448E+04 \\
 2 & 1.969207E+08 & 2.684860E+06 & 5.026387E+07 & 5.221891E+04 & 2.989959E+00 & 3.937624E+03 & 1.183492E+05 \\
 3 & 2.086133E+08 & 2.863514E+06 & 5.204397E+07 & 6.784875E+04 & 2.989958E+00 & 4.585147E+03 & 2.987256E+05 \\
 4 & 2.195342E+08 & 3.031484E+06 & 5.371028E+07 & 8.031166E+04 & 2.989955E+00 & 5.245723E+03 & 7.595382E+05 \\
 5 & 2.269594E+08 & 3.190444E+06 & 5.494672E+07 & 8.659560E+04 & 2.989951E+00 & 5.779177E+03 & 1.319066E+06 \\
 6 & 2.342317E+08 & 3.365202E+06 & 5.611659E+07 & 8.875659E+04 & 2.989945E+00 & 6.321060E+03 & 2.281208E+06 \\
 7 & 2.394346E+08 & 3.526668E+06 & 5.707119E+07 & 8.778255E+04 & 2.989937E+00 & 6.797777E+03 & 3.222215E+06 \\
 8 & 2.440928E+08 & 3.692101E+06 & 5.796298E+07 & 8.513058E+04 & 2.989924E+00 & 7.263765E+03 & 4.398030E+06 \\
 9 & 2.478168E+08 & 3.866263E+06 & 5.875260E+07 & 8.134384E+04 & 2.989906E+00 & 7.706778E+03 & 5.538496E+06 \\
10 & 2.513340E+08 & 4.048897E+06 & 5.950483E+07 & 7.718683E+04 & 2.989879E+00 & 8.141840E+03 & 6.776800E+06 \\
11 & 2.551253E+08 & 4.215980E+06 & 6.026202E+07 & 7.320287E+04 & 2.989840E+00 & 8.584714E+03 & 8.464204E+06 \\
12 & 2.589790E+08 & 4.410556E+06 & 6.096342E+07 & 6.977448E+04 & 2.989780E+00 & 9.015570E+03 & 1.072477E+07 \\
13 & 2.630434E+08 & 4.610824E+06 & 6.166960E+07 & 6.747551E+04 & 2.989682E+00 & 9.461654E+03 & 1.447887E+07 \\
14 & 2.682954E+08 & 4.819258E+06 & 6.234783E+07 & 6.636544E+04 & 2.989504E+00 & 9.909229E+03 & 1.847517E+07 \\
15 & 2.715982E+08 & 5.026276E+06 & 6.293235E+07 & 6.538550E+04 & 2.989177E+00 & 1.032697E+04 & 3.002030E+07 \\
16 & 2.781746E+08 & 5.236472E+06 & 6.348534E+07 & 6.524187E+04 & 2.988530E+00 & 1.073885E+04 & 4.387040E+07 \\
17 & 2.808189E+08 & 5.397694E+06 & 6.389052E+07 & 6.336352E+04 & 2.987350E+00 & 1.108444E+04 & 4.838104E+07 \\
18 & 2.832818E+08 & 5.603128E+06 & 6.432855E+07 & 6.284853E+04 & 2.985032E+00 & 1.144775E+04 & 5.528658E+07 \\
19 & 2.867869E+08 & 5.794352E+06 & 6.465809E+07 & 6.148713E+04 & 2.980963E+00 & 1.176355E+04 & 8.176797E+07 \\
20 & 2.898100E+08 & 6.002216E+06 & 6.496616E+07 & 6.025162E+04 & 2.973891E+00 & 1.207047E+04 & 9.513553E+07 \\
21 & 2.926001E+08 & 6.200394E+06 & 6.525625E+07 & 5.917455E+04 & 2.962824E+00 & 1.236732E+04 & 1.163203E+08 \\
22 & 2.953747E+08 & 6.541253E+06 & 6.554327E+07 & 5.878434E+04 & 2.951519E+00 & 1.266948E+04 & 1.445130E+08 \\
23 & 2.972316E+08 & 6.751874E+06 & 6.573163E+07 & 5.686887E+04 & 2.936666E+00 & 1.291685E+04 & 1.500224E+08 \\
24 & 2.999315E+08 & 6.557310E+06 & 6.592716E+07 & 5.622341E+04 & 2.845449E+00 & 1.314644E+04 & 1.755376E+08 \\
25 & 3.021500E+08 & 6.128534E+06 & 6.598145E+07 & 5.471787E+04 & 2.211516E+00 & 1.323496E+04 & 1.658657E+08 \\
\hline
\hline
\end{tabular}
\medskip\\
\end{center}
\end{table}
\end{landscape}

\begin{table}[ht]
\begin{center}
\caption{Data used to estimate the number of remaining TPs. 
We include the final total and core mass after the final TP,
as determined by the synthetic AGB program, the $\lambda$ used for the 
final TDU episodes, the total 
number of TPs (including the synthetic TPs) and the number of synthetic TPs.
The VW93 indicates the 5$\Msun$, $Z= 0.02$ model that was computed with \citet{vw93}
mass loss on the AGB whereas the R75 indicates data for the model computed
with Reimer's mass loss with $\eta = 3.5$.}
\begin{tabular}{lccccc} \hline
 \multicolumn{6}{c}{\em Z=0.02} \\ \hline
$M_{0}$ & $M_{\rm f}$ & $M_{c,f}$ & $\lambda_{\rm f}$ & No. TPs & Syn. TPs \\ \hline
2.5  & 1.1388 & 0.6630 & 0.4584 & 26 & 1 \\
3.0  & 1.3580 & 0.6819 & 0.6661 & 26 & 1 \\
3.5  & 1.2675 & 0.7181 & 0.8583 & 22 & 1 \\
4.0  & 1.1151 & 0.7916 & 0.9306 & 19 & 2 \\
5.0 (VW93)  & 1.2545 & 0.8757 & 0.9098 & 27 & 4 \\
5.0 (R75)   & 1.0458 & 0.8736 & 0.9516 & 42 & 5  \\
6.0  & 1.1984 & 0.9294 & 0.9471 & 43 & 5 \\
6.5  & 0.9918 & 0.9629 & 0.8539 & 46 & 7 \\ \hline
\multicolumn{6}{c}{\em Z=0.008} \\ \hline
2.5  & 0.9978 & 0.6626 & 0.7760 & 28 & 1  \\
3.0  & 1.0077 & 0.6934 & 0.8612 & 29 & 1  \\
3.5  & 1.4192 & 0.7662 & 0.9530 & 21 & 1  \\
4.0  & 0.9716 & 0.8367 & 0.9730 & 24 & 2  \\
5.0  & 1.1631 & 0.8860 & 0.9540 & 61 & 4  \\
6.0  & 1.1408 & 0.9480 & 0.9195 & 73 & 5  \\ \hline
\multicolumn{6}{c}{\em Z=0.004} \\ \hline
2.5  & 1.3578 & 0.6707 & 0.8165 & 29 & 1 \\ 
3.5  & 1.1274 & 0.8138 & 0.9720 & 24 & 1 \\
4.0  & 1.0533 & 0.8524 & 0.9460 & 32 & 1 \\
5.0  & 1.0138 & 0.9055 & 0.9715 & 84 & 3 \\
6.0  & 1.1859 & 0.9792 & 0.9515 & 106 & 5 \\ \hline
\multicolumn{6}{c}{\em Z=0.0001} \\ \hline
2.5  & 1.1489 & 0.7325 & 0.8207 & 31 & 1  \\
3.0  & 1.2758 & 0.8159 & 0.9691 & 40 & 1  \\
3.5  & 1.2627 & 0.8474 & 0.9505 & 59 & 1  \\
4.0  & 1.1767 & 0.8694 & 0.9188 & 76 & 1  \\
5.0 (VW93) & 1.0374 & 0.9340 & 0.9297 & 138 & 2  \\
5.0 (R75)  & 1.2558 & 0.9244 & 0.9312 & 69  & 1  \\ \hline
\end{tabular}  \label{missedTPs}
\end{center}
\end{table}

\begin{landscape}  
\begin{table}[t]
\begin{center}
\caption{Stellar yields for 3$\Msun$, $Z = 0.02$ model.}\label{yield1}
\begin{tabular}{ccrrrrrr}
\hline Isotope $i$ &  $A$ & yield & mass$(i)_{\rm lost}$ &  mass$(i)_{0}$ & $\langle X(i) \rangle$ 
& $X0(i)$ & $f$ \\
\hline
\hline
$g^{a}$    &  1 & 1.0600706E-07 & 1.2625863E-05 & 1.2519856E-05 & 5.4468778E-06 & 5.4014972E-06 & 3.6334887E-03 \\
 p         &  1 & $-$8.0847383E-02 & 1.5124202 & 1.5932676 &  6.5246773E-01 & 6.8739051E-01 & $-$2.2644492E-02 \\
\iso{3}He  &  3 & 2.2212508E-04 & 2.2214468E-04 & 1.9598588E-08 & 9.5834628E-05 & 8.4555056E-09 & 4.0543828 \\
\iso{4}He  &  4 & 6.1267972E-02 & 7.4012238E-01 & 6.7885441E-01 & 3.1929350E-01 & 2.9288119E-01 & 3.7498605E-02 \\
\iso{7}Li  &  7 & $-$1.8814418E-08 & 6.0166285E-09 & 2.4831047E-08 & 2.5956119E-09 & 1.0712970E-08 & $-$6.1567014E-01 \\
\iso{12}C  & 12 & 1.4943778E-02 & 2.2845197E-02 & 7.9014199E-03 & 9.8555638E-03 & 3.4089447E-03 & 4.6106154E-01 \\
\iso{13}C  & 13 & 1.1727007E-04 & 2.1251373E-04 & 9.5243653E-05 & 9.1679773E-05 & 4.1091393E-05 & 3.4852266E-01 \\
\iso{14}N  & 14 & 3.3946566E-03 & 5.8388105E-03 & 2.4441539E-03 & 2.5189000E-03 & 1.0544922E-03 & 3.7816754E-01 \\
\iso{15}N  & 15 & $-$4.7271697E-06 & 4.8713855E-06 & 9.5985552E-06 & 2.1015467E-06 & 4.1411477E-06 & $-$2.9458165E-01 \\
\iso{16}O  & 16 & $-$1.2392532E-03 & 2.1018269E-02 & 2.2257522E-02 & 9.0674153E-03 & 9.6026622E-03 & $-$2.4908133E-02 \\
\iso{17}O  & 17 & 5.1704166E-05 & 6.0690640E-05 & 8.9864743E-06 & 2.6182328E-05 & 3.8770745E-06 & 8.2950413E-01 \\
\iso{18}O  & 18 & $-$1.3407498E-05 & 3.6671925E-05 & 5.0079423E-05 & 1.5820502E-05 & 2.1605989E-05 & $-$1.3535389E-01 \\
\iso{19}F  & 19 & 3.8715425E-06 & 4.9463952E-06 & 1.0748529E-06 & 2.1339065E-06 & 4.6372855E-07 & 6.6291153E-01 \\
\iso{20}Ne & 20 & $-$1.1501834E-06 & 4.1906834E-03 & 4.1918335E-03 & 1.8078875E-03 & 1.8085014E-03 & $-$1.4747151E-04 \\
\iso{21}Ne & 21 & 4.0954365E-07 & 1.1120977E-05 & 1.0711433E-05 & 4.7976600E-06 & 4.6212808E-06 & 1.6267121E-02 \\
\iso{22}Ne & 22 & 1.9907814E-03 & 2.3274263E-03 & 3.3664503E-04 & 1.0040666E-03 & 1.4524026E-04 & 8.3967549E-01 \\
\iso{23}Na & 23 & 7.5442040E-05 & 1.6396206E-04 & 8.8520021E-05 & 7.0734277E-05 & 3.8190585E-05 & 2.6767361E-01 \\
\iso{24}Mg & 24 & $-$8.9756213E-07 & 1.3679401E-03 & 1.3688377E-03 & 5.9013808E-04 & 5.9056369E-04 & $-$3.1310008E-04 \\
\iso{25}Mg & 25 & 2.4309760E-05 & 2.0399288E-04 & 1.7968312E-04 & 8.8003828E-05 & 7.7521494E-05 & 5.5079415E-02 \\
\iso{26}Mg & 26 & 1.1070690E-05 & 2.1723099E-04 & 2.0616030E-04 & 9.3714836E-05 & 8.8944660E-05 & 2.2688469E-02 \\
\iso{26}Al & 26 & 1.1710374E-07 & 1.1710374E-07 & 0.0000000E+00 & 5.0519297E-08 & 0.0000000E+00 & 0.0000000E+00 \\
\iso{27}Al & 27 & 1.2958917E-06 & 1.5529012E-04 & 1.5399423E-04 & 6.6993147E-05 & 6.6438413E-05 & 3.6111036E-03 \\
\iso{28}Si & 28 & $-$1.6319100E-06 & 1.7316726E-03 & 1.7333045E-03 & 7.4705458E-04 & 7.4780727E-04 & $-$4.3735650E-04 \\
\iso{29}Si & 29 & 1.5787009E-06 & 9.2347655E-05 & 9.0768954E-05 & 3.9839368E-05 & 3.9160856E-05 & 7.4602445E-03 \\
\iso{30}Si & 30 & 2.4255860E-07 & 6.2702733E-05 & 6.2460174E-05 & 2.7050359E-05 & 2.6947471E-05 & 1.6549942E-03 \\
\iso{31}P  & 31 & 2.0692660E-06 & 2.3703114E-05 & 2.1633849E-05 & 1.0225674E-05 & 9.3335875E-06 & 3.9643381E-02 \\
\iso{32}S  & 32 & $-$2.1749875E-06 & 1.0513477E-03 & 1.0535227E-03 & 4.5355811E-04 & 4.5452599E-04 & $-$9.2578813E-04 \\
\iso{33}S  & 33 & 4.0168288E-07 & 8.9708283E-06 & 8.5691454E-06 & 3.8700723E-06 & 3.6970248E-06 & 1.9866738E-02 \\
\iso{34}S$^{b}$ & 34 & $-$1.7065395E-07 & 4.9333787E-05 & 4.9504441E-05 & 2.1282909E-05 &  2.1357921E-05 & $-$1.5279896E-03 \\
\iso{56}Fe & 56 & $-$2.1018088E-05 & 3.0820514E-03 & 3.1030695E-03 & 1.3296166E-03 & 1.3387711E-03 & $-$2.9798900E-03 \\
\iso{57}Fe & 57 & 1.3097335E-05 & 8.8821595E-05 & 7.5724260E-05 & 3.8318201E-05 & 3.2670057E-05 & 6.79255233E-02 \\
\iso{58}Fe & 58 & 7.2211260E-06 & 1.7037621E-05 & 9.8164946E-06 & 7.3501383E-06 & 4.2351739E-06 & 2.3942426E-01 \\
\iso{60}Fe & 60 & 3.9827793E-08 & 3.9827793E-08 & 0.0000000E+00 & 1.7181964E-08 & 0.0000000E+00 & 0.0000000E+00 \\
\iso{59}Co & 59 & 1.4924935E-06 & 1.0411269E-05 & 8.9187752E-06 & 4.4914877E-06 & 3.8478670E-06 & 6.7170151E-02 \\
\iso{58}Ni & 58 & $-$2.2356980E-06 & 1.2900950E-04 & 1.3124519E-04 &  5.5655517E-05 & 5.6623696E-05 & $-$7.4899960E-03 \\
\iso{60}Ni & 60 & 7.0899841E-07 & 5.2735955E-05 & 5.2026957E-05 & 2.2750626E-05 & 2.2446222E-05 & 5.8501000E-03 \\
\iso{61}Ni & 61 & 8.3594359E-06 & 1.0636427E-05 & 2.2769907E-06 & 4.5886222E-06 & 9.8237228E-07 & 6.6940618E-01 \\
\iso{62}Ni & 62 & $-$7.3240267E-06 & 5.0168190E-08 & 7.3741949E-06 & 2.1642876E-08 & 3.1814818E-06 & $-$2.1673145E+00 \\
\hline
\hline
\end{tabular}
\medskip\\
$^a$ $g$ represents the abundance of species from \iso{64}Ni to Bi; an increase in $g$ 
indicates that neutron-captures have occurred beyond the end of the network.\\
$^b$ \iso{34}S includes the abundance of species between \iso{34}S and Mn.\\
\end{center}
\end{table}
\end{landscape}

\begin{table}[t]
\begin{center}
\caption{PN yields 3$\Msun$, $Z = 0.02$ model. The first entry is the total mass ejected
during the final two TPs, 1.502\,$\Msun$ in this case.}\label{pnyield}
\begin{tabular}{ccrrrrr}
\hline Isotope $i$ &  $A$ & yield & mass$(i)_{\rm lost}$ & $X0(i)$  &  $\langle X(i) \rangle$ & $N(i)/N(H)$ \\
\hline
\hline
    g       &   1 & 9.4299139E-08 & 8.2076585E-06 & 5.4014972E-06 & 5.4642769E-06 & 8.3678888E-06 \\
    p       &   1 & $-$5.1648498E-02 & 9.8085165E-01 & 6.8739051E-01 & 6.5300536E-01 & 1.0000000E+00 \\
\iso{3}He   &   3 & 1.4400885E-04 & 1.4402156E-04 & 8.4555056E-09 & 9.5882846E-05 & 4.8944388E-05 \\
\iso{4}He   &   4 & 3.9245605E-02 & 4.7917002E-01 & 2.9288119E-01 & 3.1900910E-01 & 1.2213111E-01 \\
\iso{7}Li   &   7 & $-$1.2191149E-08 & 3.9003480E-09 & 1.0712970E-08 & 2.5966702E-09 & 5.6807015E-10 \\
\iso{12}C   &  12 & 9.4264597E-03 & 1.4546891E-02 & 3.4089447E-03 & 9.6846428E-03 & 1.2359064E-03 \\
\iso{13}C   &  13 & 7.6098098E-05 & 1.3781973E-04 & 4.1091393E-05 & 9.1753966E-05 & 1.0808482E-05 \\
\iso{14}N   &  14 & 2.2036042E-03 & 3.7875122E-03 & 1.0544922E-03 & 2.5215494E-03 & 2.7581805E-04 \\
\iso{15}N   &  15 & $-$3.0603858E-06 & 3.1598568E-06 & 4.1411477E-06 & 2.1036856E-06 & 2.1476960E-07 \\
\iso{16}O   &  16 & $-$7.9538208E-04 & 1.3628369E-02 & 9.6026622E-03 & 9.0731336E-03 & 8.6840155E-04 \\
\iso{17}O   &  17 & 3.3530909E-05 & 3.9354498E-05 & 3.8770745E-06 & 2.6200392E-05 & 2.3601635E-06 \\
\iso{18}O   &  18 & $-$8.6711771E-06 & 2.3782261E-05 & 2.1605989E-05 & 1.5833122E-05 & 1.3470301E-06 \\
\iso{19}F   &  19 & 2.4399674E-06 & 3.1365144E-06 & 4.6372855E-07 & 2.0881453E-06 & 1.6830241E-07 \\
\iso{20}Ne  &  20 & $-$8.7078661E-07 & 2.7156025E-03 & 1.8085014E-03 & 1.8079218E-03 & 1.3843084E-04 \\
\iso{21}Ne  &  21 & 2.5139525E-07 & 7.1928253E-06 & 4.6212808E-06 & 4.7886483E-06 & 3.4920211E-07 \\
\iso{22}Ne  &  22 & 1.2537986E-03 & 1.4719579E-03 & 1.4524026E-04 & 9.7996101E-04 & 6.8213347E-05 \\
\iso{23}Na  &  23 & 4.7911446E-05 & 1.0527590E-04 & 3.8190585E-05 & 7.0087794E-05 & 4.6665705E-06 \\
\iso{24}Mg  &  24 & $-$5.7858415E-07 & 8.8648207E-04 & 5.9056369E-04 & 5.9017848E-04 & 3.7657839E-05 \\
\iso{25}Mg  &  25 & 1.4788740E-05 & 1.3123048E-04 & 7.7521494E-05 & 8.7367145E-05 & 5.3516951E-06 \\
\iso{26}Mg  &  26 & 6.7903602E-06 & 1.4039036E-04 & 8.8944660E-05 & 9.3465365E-05 & 5.5050414E-06 \\
\iso{26}Al  &  26 & 7.8603932E-08 & 7.8603932E-08 & 0.0000000E+00 & 5.2330840E-08 & 3.0822480E-09 \\
\iso{27}Al  &  27 & 8.2996121E-07 & 1.0062428E-04 & 6.6438413E-05 & 6.6990964E-05 & 3.7995808E-06 \\
\iso{28}Si  &  28 & $-$1.0048971E-06 & 1.1222448E-03 & 7.4780727E-04 & 7.4713834E-04 & 4.0862618E-05 \\
\iso{29}Si  &  29 & 9.7531665E-07 & 5.9797174E-05 & 3.9160856E-05 & 3.9810177E-05 & 2.1022256E-06 \\
\iso{30}Si  &  30 & 1.4533725E-07 & 4.0621992E-05 & 2.6947471E-05 & 2.7044231E-05 & 1.3805007E-06 \\
\iso{31}P   &  31 & 1.2756063E-06 & 1.5295192E-05 & 9.3335875E-06 & 1.0182827E-05 & 5.0302543E-07 \\
\iso{32}S   &  32 & $-$1.3433164E-06 & 6.8138086E-04 & 4.5452599E-04 & 4.5363166E-04 & 2.1708838E-05 \\
\iso{33}S   &  33 & 2.5333611E-07 & 5.8064797E-06 & 3.6970248E-06 & 3.8656840E-06 & 1.7938892E-07 \\
\iso{34}S   &  34 & $-$1.0619988E-07 & 3.1974625E-05 & 2.1357921E-05 & 2.1287216E-05 & 9.5878931E-07 \\
\iso{56}Fe  &  56 & $-$1.3052253E-05 & 1.9978590E-03 & 1.3387711E-03 & 1.3300815E-03 & 3.6372523E-05 \\
\iso{57}Fe  &  57 & 8.2095721E-06 & 5.7281879E-05 & 3.2670057E-05 & 3.8135608E-05 & 1.0245640E-06 \\
\iso{58}Fe  &  58 & 4.4331669E-06 & 1.0794642E-05 & 4.2351739E-06 & 7.1865697E-06 & 1.8974787E-07 \\
\iso{60}Fe  &  60 & 2.0917762E-08 & 2.0917762E-08 & 0.0000000E+00 & 1.3926072E-08 & 3.5543535E-10 \\
\iso{59}Co  &  59 & 9.3833069E-07 & 6.7180481E-06 & 3.8478670E-06 & 4.4725639E-06 & 1.1608812E-07 \\
\iso{58}Ni  &  58 & $-$1.3969257E-06 & 8.3655119E-05 & 5.6623696E-05 & 5.5693683E-05 & 1.4704869E-06 \\
\iso{60}Ni  &  60 & 4.3515684E-07 & 3.4150675E-05 & 2.2446222E-05 & 2.2735929E-05 & 5.8028951E-07 \\
\iso{61}Ni  &  61 & 5.3855811E-06 & 6.8611607E-06 & 9.8237228E-07 & 4.5678416E-06 & 1.1467385E-07 \\
\iso{62}Ni  &  62 & $-$4.7543513E-06 & 2.4417719E-08 & 3.1814818E-06 & 1.6256180E-08 & 4.0152268E-10 \\
\hline
\hline
\end{tabular}
\end{center}
\end{table}

\end{document}